\documentclass{LMCS}

\usepackage{cite,url,hyperref}
\usepackage{bbm,stmaryrd,mathbbol}
\usepackage[neverdecrease]{paralist}				
\usepackage{picins,graphicx,gastex}
\usepackage{enumerate}
\allowdisplaybreaks

\newenvironment{definition}{\begin{defi}}{\end{defi}}
\newenvironment{lemma}{\begin{lem}}{\end{lem}}
\newenvironment{corollary}{\begin{cor}}{\end{cor}}
\newenvironment{proposition}{\begin{prop}}{\end{prop}}
\newenvironment{theorem}{\begin{thm}}{\end{thm}}
\newenvironment{remark}{\begin{rem}}{\end{rem}}
\newenvironment{example}{\begin{exa}}{\end{exa}}

\DeclareSymbolFont{rsfs}{U}{rsfs}{m}{n}
\DeclareSymbolFontAlphabet{\mathrsfs}{rsfs}

\newcommand{\A}    	{\ensuremath{\mathcal{A}}}
\newcommand{\Bo}    	{\ensuremath{\mathbb{B}}}
\newcommand{\C}    	{\ensuremath{\mathcal{C}}}
\newcommand{\D}    	{\ensuremath{\mathcal{D}}}
\newcommand{\K}    	{\ensuremath{\mathbb{K}}}
\newcommand{\N} 	{\ensuremath{\mathbb{N}}}
\newcommand{\PowS}    	{\ensuremath{\mathrsfs{P}}}
\newcommand{\R}    	{\ensuremath{\mathbb{R}}}        
\newcommand{\V}	        {\ensuremath{\mathcal{V}}}
\newcommand{\La}        {\ensuremath{\mathrsfs{L}}}
\newcommand{\cX}	{\ensuremath{\mathcal{X}}}
\newcommand{\Z}    	{\ensuremath{\mathbb{Z}}}

\newcommand{\PA}    	{\ensuremath{\mathcal{P}}}
\newcommand{\HS}  	{\ensuremath{\mathcal{H}}}
\newcommand{\VS}	{\ensuremath{\mathcal{V}}}
\newcommand{\behave}[1]	{\ensuremath{\parallel \!\! {#1} \!\! \parallel}}
\newcommand{\seman}[1]	{\ensuremath{\llbracket {#1} \rrbracket}}
\DeclareMathOperator{\Lab}	{Lab}
\DeclareMathOperator{\lab}      {lab}

\DeclareMathOperator{\frst}    	{Frst}

\DeclareMathOperator{\next}	{next}
\DeclareMathOperator{\call}	{call}
\DeclareMathOperator{\return}	{return}
\DeclareMathOperator{\TXT}	{TXT}
\DeclareMathOperator{\NW}	{NW}
\DeclareMathOperator{\Free}	{Free}
\DeclareMathOperator{\Der}      {Der}
\DeclareMathOperator{\Nest}     {Nest}
\DeclareMathOperator{\TDO}      {TDO}
\DeclareMathOperator{\weight}   {wgt}
\DeclareMathOperator{\initial}  {init}
\DeclareMathOperator{\final}    {fin}
\DeclareMathOperator{\dom}	{dom}
\DeclareMathOperator{\supp}	{supp}

\DeclareMathOperator{\MSO}    	{MSO}
\DeclareMathOperator{\RMSO}    	{RMSO}

\DeclareMathOperator{\FO}    	{FO}

\DeclareMathOperator{\sRFO}    	{sRFO}
\DeclareMathOperator{\EMSO}    	{EMSO}
\DeclareMathOperator{\aUMSO}	{aUMSO}
\DeclareMathOperator{\wUMSO}	{wUMSO}
\DeclareMathOperator{\sRMSO}	{sRMSO}
\DeclareMathOperator{\swRMSO}	{swRMSO}
\DeclareMathOperator{\sREMSO}	{sREMSO}
\newcommand{\deftrans}	{\ensuremath{\mathbf{def}}}
\newcommand{\muop}      {\ensuremath{\mu_{\text{op}}}}
\newcommand{\mucl}      {\ensuremath{\mu_{\text{cl}}}}
\newcommand{\cha}[1]    {\ensuremath{\mathbb{1}_{#1}}}
\newcommand{\disjoint}  {\ensuremath{\uplus}}
\newcommand{\bottom}    {\ensuremath{\bot}}
\newcommand{\impl}      {\ensuremath{\xrightarrow{+}}}
\renewcommand{\phi}	{\ensuremath{\varphi}}
\renewcommand{\theta}	{\ensuremath{\vartheta}}
\renewcommand{\epsilon}	{\ensuremath{\varepsilon}}
\newcommand{\sigtxt}[1] {\ensuremath{#1,\leq_1,\leq_2}}
\newcommand{\signw}[1]  {\ensuremath{#1,\leq,\nu}}
\newcommand{\deltaint}  {\delta_{\text{int}}} 
\newcommand{\deltacall} {\delta_{\text{call}}} 
\newcommand{\deltaret}  {\delta_{\text{ret}} }

\newcommand{\eop}     {}

\def\doi{6 (1:5) 2010}
\lmcsheading%
{\doi}
{1--34}
{}
{}
{Dec.~\phantom08, 2008}
{Feb.~19, 2010}
{}   

\begin{document}

\title[Weighted Logics for Nested Words and
  Algebraic Formal Power Series]{Weighted Logics for Nested Words and \\
  Algebraic Formal Power Series} 
\date{}
\author[C.~Mathissen]{Christian Mathissen}
\address{ Institut f{\"u}r Informatik,
  Universit{\"at} Leipzig,
  04009 Leipzig, Germany}
\email{mathissen@informatik.uni-leipzig.de}
\thanks{Supported by the Graduiertenkolleg 446 of the German Research
  Foundation (DFG)}
\titlecomment{
%
 An extended abstract of this paper appeared in the proceedings of the
 35th ICALP, Reykjavik, 2008~\cite{Mat08}.
}
\begin{abstract}  
  \noindent Nested words, a model for recursive programs proposed by
  Alur and Madhusudan, have recently gained much interest. In this
  paper we introduce quantitative extensions and study nested word
  series which assign to nested words elements of a semiring.  We show
  that regular nested word series coincide with series definable in
  weighted logics as introduced by Droste and Gastin.  For this we
  establish a connection between nested words and the free
  bisemigroup.  Applying our result, we obtain characterizations of
  algebraic formal power series in terms of weighted logics. This
  generalizes results of Lautemann, Schwentick and Th{\'e}rien on
  context-free languages.
\end{abstract}
\maketitle

\section{Introduction}
\label{sec:intro}

Model checking of finite state systems has become an established
method for automatic hardware and software verification and led to
numerous verification programs used in industrial application.  In
order to verify recursive programs it is necessary to model them as
pushdown systems rather than finite automata.  This has motivated Alur
and Madhusudan \cite{AluMad04b,AluMad06} to define regular nested word
languages and visibly pushdown languages. The latter is a proper
subclass of the context-free languages and exceeds the regular
languages. Both classes are closely related.  Nested words on the one
hand have a linear sequential structure and on the other hand have a
hierarchical structure.  This way they may also be used to model
linguistic data as well as semistructured data such as XML documents.
Nested words and visibly pushdown languages gained much interest and
set a starting point for a new research field (see e.g.
\cite{Aluetal05,CheWal07,Aluetal07} among many others).

The goal of this paper is: 1. to introduce a quantitative automaton
model and a quantitative logic for nested words that are equally
expressive, 2. to establish a connection between nested words and
alternating texts, a graph representation of the free bisemigroup
which is an object studied by {\'E}sik and N{\'e}meth \cite{EsiNem04}
and Hashiguchi et al.~\cite{Hasetal00,Hasetal02,Hasetal03}, 3. to give
a characterization of the important class of algebraic formal power
series by means of weighted logics.

In order to model quantitative aspects, extensions of existing models
such as weighted automata were investigated.  There, transitions of
automata additionally carry a weight which can be of very different
nature (e.g. counting, probabilities, etc.). In fact, weighted
automata have found many different applications e.g. in image
processing~\cite{CulKar93}, in speech recognition~\cite{Moh97} or as a
model for probabilistic systems~\cite{BaiGro05,Bai08}. In this paper
we introduce and investigate weighted nested word automata which may
serve as a quantitative model for sequential programs with recursive
procedure calls.  Due to the fact that we define them over arbitrary
semirings, they are very flexible and can model, for example,
probabilistic or stochastic programs of recursive nature as well as
quantitative database queries.

Since weighted nested word automata and weighted pushdown automata are
closely related, one should also mention that weighted pushdown
systems have been applied to data flow analysis (see
e.g. \cite{RepLalKid07,Repsetal05}). There, however, the emphasize
lies on the (weighted) configuration graph of the system which is used
to model the state space of a program. Weights are incorporated in
order to model, for example, the data of the program. In
\cite{RepLalKid07,Repsetal05} weighted versions of reachability
problems in such graphs were considered.

In this paper we are interested in the semantics of a weighted
automaton given as a mapping which assigns a value to each nested
word.  As the first main result of this paper we characterize the
expressiveness of weighted nested word automata using weighted logics,
generalizing a result of Alur and Madhusudan.  Weighted logics were
introduced by Droste and Gastin \cite{DroGas07}.  They enriched the
classical language of monadic second-order logic with values from a
semiring in order to add quantitative expressiveness.
This way one may now e.g. express \emph{how often} a certain property
holds, \emph{how much} execution time a process needs or \emph{how}
reliable it is.  The result of Droste and Gastin has been extended to
infinite words, (infinite) trees, texts, pictures and
traces~\cite{DroRah06,DroVog06a,Mae06,Mat07,Mei06,Rah07}. We note,
moreover, that a restriction of {\L}ukasiewicz multi-valued logic
coincides with this weighted logics~\cite{Sch07}.

In order to prove our result mentioned above we establish a new
connection between alternating texts and nested words and reduce the
result to an analogous one for alternating texts. The class of
alternating texts, introduced by Ehrenfeucht and
Rozenberg~\cite{EhrRoz93}, forms the free bisemigroup which was also
investigated by Hashiguchi et
al.~\cite{Hasetal00,Hasetal02,Hasetal03}.  Moreover, a language theory
for series-parallel-biposets, a different representation of the free
bisemigroup, was developed by {\'E}sik and N{\'e}meth \cite{EsiNem04}.
Besides the author's opinion that a reduction to a previously known
result is mathematically more elegant than e.g. a structural
induction, the approach admits the advantage that it gives insight
into relationships and similarities between different structures
considered in the literature and therefore offers benefits. For
example, decidability results for the emptiness and equivalence
problem come almost for free as a corollary.  Note that this extends
the classical satisfiability problem for monadic second order logic,
which is one motivation of transforming formulas in automata.

Furthermore, we can use the connection again in this paper to obtain a
new characterization of algebraic formal power series.  The latter
form an important generalization of context-free languages.  Algebraic
formal power series were considered initially already by Chomsky and
Sch\"utzenberger \cite{ChoSch63} and have since been intensively
studied by Kuich and others.  For a survey see \cite{Kui97} or
\cite{KuiSal86}. Using projections of nested word series and applying
the logical characterization of weighted nested word automata, we are
able to give a characterization of algebraic formal power series in
terms of weighted logics, generalizing a result of Lautemann,
Schwentick and Th{\'e}rien \cite{Lauetal94} on context-free
languages. The connection between alternating texts and nested words
is then used to also generalize a second characterization
of~\cite{Lauetal94}, thereby giving a different proof also for the
result of Lautemann, Schwentick and Th{\'e}rien.

The paper is organized as follows. In Section~\ref{sec:nw} we
introduce nested words, weighted automata for nested words and give an
example for them. In Section~\ref{sec:mso} we introduce weighted
logics for nested words, introduce different fragments of the latter
and state the first main result, the characterization of regular
nested word series in terms of weighted logics. In
Section~\ref{sec:spb} we introduce alternating texts, a graph
representation of the free bisemigroup and define a weighted version
of {\'E}sik and N{\'e}meth's parenthesizing automata operating over
elements of the free bisemigroup. Next, in
Section~\ref{sec:interpr-nest-words}, we define an embedding of nested
words into alternating texts and show that we can translate weighted
formulae as well as automata back and forth with respect to this
embedding. This gives the proof of the first main result. After that,
in Section~\ref{cha:algebraic}, we apply the result and obtain
characterizations of algebraic formal power series in terms of
weighted logics.

An extended abstract of this paper appeared as~\cite{Mat08}. This
paper differs from it in the following way. First, full proofs are
included. Second, the first main result, the logical characterization
of regular nested word series, has been extended and it is shown that
an existential fragment of weighted logics suffices to characterize
weighted automata over nested words. Third, rather than translating
nested words to sp-biposets, the graph representation of the free
bisemigroup used by {\'E}sik and N{\'e}meth~\cite{EsiNem04}, we
translate it to alternating texts, a different representation. This
admits the advantage that we can more easily obtain a second
characterization of algebraic formal power series in terms of weighted
logics. This second characterization, which we include here in full
length, was only sketched in the concluding remarks of \cite{Mat08}
and gives the fourth main difference.

\section{Weighted Automata on Nested Words}
\label{sec:nw}

In this section we recall the notion of nested words which was
introduced by Alur and Madhusudan~\cite{AluMad06} and we define
weighted automata for them. Let $\Delta$ be a finite alphabet and let
$\Delta^+$ be the free semigroup of finite but non-empty words. Let
$w=a_1\ldots a_n \in \Delta^+$. The length of $w$ is $|w|=n$.  A
\emph{nesting relation $\nu$ of width} $n$ ($n \in \N$) is a binary
relation on $[n]=\{1, \ldots ,n\}$ such that for all $1 \leq i,j\leq
n$:
\begin{enumerate}[(1)]
\item if \emph{$\nu(i,j)$,} then \emph{$i <j$},
\item if \emph{$\nu(i,j)$ and $\nu(i,j')$,} then \emph{$j=j'$} and if
  \emph{$\nu(i,j)$ and $\nu(i',j)$}, then \emph{$i=i'$},
\item if \emph{$\nu(i,j)$ and $\nu(i',j')$ and $i < i'$} then either
  \emph{$j < i'$ or $j' < j$}. 
\end{enumerate}
If $\nu(i,j)$, we say $i$ is a \emph{call position} and $j$ is a
\emph{return position.} Any $1\leq i \leq n$ which is neither a call
nor a return position is called an \emph{internal position}. We
collect all nesting relations of width $n$ in $\Nest_n$.
\begin{definition}[Alur \& Madhusudan~\cite{AluMad06}]\label{def:nw}
  A \emph{nested word} (over $\Delta$) is a pair $(w,\nu)$ such that
  $w \in \Delta^+$ and $\nu$ is a \emph{nesting relation of width
    $|w|$.}
\end{definition}
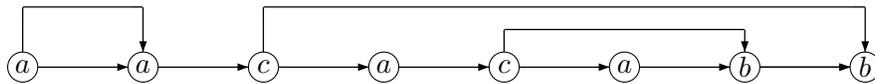
\begin{figure}[h!] 
  \begin{center}
  \begin{picture}(112,10)(0,8) \nullfont
    \node[Nh=4,Nw=4](n0)(0,10){$a$}
    \node[Nh=4,Nw=4](n1)(16,10){$a$}
    \node[Nw=4,Nh=4](n2)(32,10){$c$}
    \node[Nw=4,Nh=4](n3)(48,10){$a$}
    \node[Nh=4,Nw=4](n4)(64,10){$c$}
    \node[Nh=4,Nw=4](n5)(80,10){$a$}
    \node[Nw=4,Nh=4](n6)(96,10){$b$}    
    \node[Nw=4,Nh=4](n7)(112,10){$b$}
    \node[Nh=0,Nw=0,Nframe=n](h0)(0,18){}
    \node[Nh=0,Nw=0,Nframe=n](h1)(16,18){}
    \node[Nw=0,Nframe=n,Nh=0](h2)(32,18){}
    \node[Nw=0,Nframe=n,Nh=0](h3)(48,18){}
    \node[Nh=0,Nw=0,Nframe=n](h4)(64,15){}
    \node[Nh=0,Nw=0,Nframe=n](h5)(80,15){}
    \node[Nw=0,Nframe=n,Nh=0](h6)(96,15){}
    \node[Nw=0,Nframe=n,Nh=0](h7)(112,18){}
    \drawedge[linewidth=0.17](n0,n1){}
    \drawedge[linewidth=0.17](n1,n2){}
    \drawedge[linewidth=0.17](n2,n3){}
    \drawedge[linewidth=0.17](n3,n4){}
    \drawedge[linewidth=0.17](n4,n5){}
    \drawedge[linewidth=0.17](n5,n6){}
    \drawedge[linewidth=0.17](n6,n7){}
    \drawedge[linewidth=0.17](n6,n7){}
    \drawedge[linewidth=0.17,AHLength=0](n0,h0){}
    \drawedge[linewidth=0.17,AHLength=0](h0,h1){}
    \drawedge[linewidth=0.17](h1,n1){}
    \drawedge[linewidth=0.17,AHLength=0](n2,h2){}
    \drawedge[linewidth=0.17,AHLength=0](h2,h7){}
    \drawedge[linewidth=0.17](h7,n7){}
    \drawedge[linewidth=0.17,AHLength=0](n4,h4){}
    \drawedge[linewidth=0.17,AHLength=0](h4,h6){}
    \drawedge[linewidth=0.17](h6,n6){}
  \end{picture}
\end{center}
\caption{A visualization of the nested word $(aacacabb,\{(1,2),(3,8),(5,7)\})$}
  \label{fig:nw}
\end{figure}
We collect all nested words over $\Delta$ in $\NW(\Delta)$.  Let
$nw=(w,\nu) \in \NW(\Delta)$ where $w=a_1 \ldots a_n$. The
\emph{factor} $nw[i,j]$ for $1 \leq i \leq j\leq n$ is the restriction
of $nw $ to the positions from $i$ to $j$; more formally $nw[i,j]=(a_i
\ldots a_j, \nu[i,j])$ where $\nu[i,j]=\{(k,\ell) \mid 1 \leq k,\ell \leq
j-i+1, (k+i-1,\ell+i-1) \in \nu\}$. Furthermore, we say a pair $(k,\ell) \in
\nu$ is a \emph{surface arch} of $nw$ if there does not exist $(k',\ell')
\in \nu$ with $k'<k<\ell<\ell'$.

Nested words have been introduced in order to model executions of
recursive programs as well as nested data structures such as XML
documents. Here, we model quantitative behavior of systems or
documents such as the runtime or the probability of an execution of a
randomized program, or the number of occurrences of a certain type of
entry in an XML document.  We do this by assigning to a nested word a
quantity expressing, for example, the runtime or the probability or
the number of entries.
\begin{example}\label{ex:nw_appl} \hfill
  \begin{enumerate}[(1)]
  \item As Alur and Madhusudan point out, XML documents or bibtex
    databases can naturally be modeled as nested words, where the
    nesting relation captures open and close tags \cite{AluMad06}.
    Suppose we model bibtex databases as nested words. Then we may
    assign to a nested word e.g. the number of technical reports it
    stores.
  \item 
        \parpic[r]{ 
        \parbox{0.32\linewidth}{\mbox{}\\\texttt{{\bf proc} bar()\{ \newline
          \mbox{}\hspace{0em}read(x); \newline
          \mbox{}\hspace{0em}flip(Y);{\bf if}(Y==head) \newline
          \mbox{}\hspace{3em}beep;\newline
          \mbox{}\hspace{2em}{\bf else} \\
          \mbox{}\hspace{3em}bar(); \\
          \mbox{}\hspace{0em}flip(Y);{\bf while}(Y==head) \\
          \mbox{}\hspace{3em}write(x); \\
          \mbox{}\hspace{3em}flip(Y); \\
          \mbox{}\hspace{0em}{\bf exit};\}}
      }}
    \hspace{-1.1em}Probabilistic automata have been used to model systems with
    uncertainty, such as communication systems over lossy channels, to
    model fault-tolerant systems or to model randomized programs. 
    Consider the randomized recursive pseudo-procedure \texttt{bar}
    where \texttt{flip(Y)} means flipping a fair coin \texttt{Y}.
    Consider furthermore the alphabet $\Delta=\{r,w,b,call,ret\}$ of
    atomic events which stand for read, write, beep, call and return.
    Now, an execution of \texttt{bar} could be as follows:
    \texttt{read(x)}, flip a coin and see tail, call recursively
    \texttt{bar}, \texttt{read(x)}, flip a coin and see head,
    \texttt{beep}, flip a coin and see tail, return from the recursive
    call, flip a coin and see head, \texttt{write(x)}, flip a coin and
    see head, \texttt{write(x)}, flip a coin and see tail, exit the
    program.  Then the nested word $nw=(w,\nu)$ defined by
    $w=r.call.r.b.ret.w.w.ret$ and $\nu=\{(2,5)\}$ models this
    execution of \texttt{bar} where $\nu$ encodes the recursive call
    of \texttt{bar}. We calculate the probability of the execution by
    multiplying the probability of each atomic action (probability
    $1/2$ for those actions that depend on a coin flip), i.e. 
    $1 \cdot 1/2 \cdot 1 \cdot 1/2 \cdot 1/2 \cdot 1/2 \cdot 1/2 \cdot
    1/2 = 1/64$. We will model \texttt{bar} using a weighted nested
    word automaton in Example~\ref{ex:WNWA_appl}, below.
  \end{enumerate}
\end{example}

\noindent To be as flexible as possible, we take the quantities we
assign to a nested word from a commutative semiring. A
\emph{commutative semiring} $\K$ is an algebraic structure $(\K,
+,\cdot, 0,1)$ such that $(\K,+,0)$ and $(\K,\cdot,1)$ are commutative
monoids, multiplication distributes over addition and $0$ is
absorbing, i.e. $0 \cdot k = k \cdot 0 = 0$ for all $k \in \K$.  For
example the natural numbers $(\N,+,\cdot,0,1)$ form a commutative
semiring. Other important examples are also the tropical semiring $(\Z
\cup\{\infty\},\min,+,\infty,0)$ and the arctic or max-plus semiring
$(\Z \cup \{-\infty\},\max,+, -\infty,0)$ which have been used to
model real-time systems or discrete event systems.  These semirings
possess the property that any finitely generated submonoid of
$(\K,+,0)$ is finite. Such semirings are called \emph{additively
  locally finite}.  Another important example of an additively locally
finite semiring is the probabilistic semiring
$([0,1],\max,\cdot,0,1)$.  We call a semiring \emph{locally finite} if
any finitely generated subsemiring is finite.  Examples include any
Boolean algebra such as the trivial Boolean algebra
$\Bo=(\{0,1\},\lor,\wedge,0,1)$ as well as $(\R _+ \cup \{\infty \},
\max, \min, 0, \infty)$ and the fuzzy semiring
$([0,1],\max,\min,0,1)$.

\emph{In the following let  $\K$ be a
  commutative semiring such that $0 \ne 1$.} 

\begin{definition}\label{def:WNWA}
  A \emph{weighted nested word automaton} (WNWA for short) is a quadruple $\A=
  (Q,\iota,\delta,\kappa)$ where $\delta=(\deltacall,\deltaint,\deltaret)$ such that 
  \begin{enumerate}[(1)]
  \item $Q$ is a finite set of states,
  \item $\deltacall,\deltaint: Q \times \Delta \times Q \to \K$ are the
    \emph{call} and \emph{internal transition functions},
  \item $\deltaret: Q \times Q \times \Delta \times Q \to \K$ is the
    \emph{return transition function},
  \item $\iota, \kappa: Q \to \K$ are the \emph{initial} and
    \emph{final distribution}. 
  \end{enumerate}
\end{definition}

\noindent A \emph{run} of $\A$ on $nw=(a_1 \ldots a_n, \nu)$ is a
\emph{sequence of states} $r=(q_0, \ldots, q_n)$; we also write $r:
q_0 \stackrel{nw}{\to} q_n$. The \emph{weight} of $r$ \emph{at
  position} $1 \leq j \leq n$ is given by
\begin{align*}
  \weight_{\A}(r,j) =
  \begin{cases}
    \deltacall(q_{j-1}, a_j, q_j) & \text{if $\nu(j,i)$ for some
      $j<i\leq n$}\\
    \deltaint(q_{j-1}, a_j, q_j) & \text{if $j$ is an internal
      position} \\
    \deltaret(q_{j-1},q_{i-1},a_j,q_j) & \text{if $\nu(i,j)$ for some
      $1\leq i<j$}.
  \end{cases}
\end{align*}
Now, the \emph{weight} of $r$ is $\weight_{\A}(r) =\prod_{1 \leq j
  \leq n} \weight_{\A}(r,j)$ and the \emph{behavior}
$\behave{\A}:\NW(\Delta) \to \K$ of $\A$ is defined by
$$
\behave{\A}(nw) = \sum_{q_0,q_n \in Q} \iota(q_0) \cdot \sum_{r: q_0
  \stackrel{nw}{\to} q_n} \weight_{\A}(r) \cdot \kappa(q_n).
$$

A function $S: \NW(\Delta) \to \K$ is called a \emph{nested word
  series}. As for formal power series we write $(S,nw)$ for $S(nw)$.
We define the \emph{scalar multiplication} $.$ and the \emph{sum} $+$
pointwise, i.e. for $k \in \K$ and any two nested word series
$S_1,S_2$ we let $(k. S_1,nw) = k\cdot (S_1,nw)$ and $(S_1 +S_2,nw) =
(S_1,nw)+(S_2,nw)$ for all $nw \in \NW(\Delta)$.  For $L \subseteq
\NW(\Delta)$ let $\cha{L}$ be the \emph{characteristic series} of $L$,
i.e. the series that assumes $1$ for all $nw \in L$ and $0$ otherwise.
A nested word series $S$ is \emph{regular} if there is a WNWA $\A$
such that $\behave{\A} = S$.  For $\K = \Bo$, i.e. when
$\deltacall,\deltaint$ and $\deltaret$ are subsets of $Q \times \Delta
\times Q$ and $Q \times Q \times \Delta \times Q$, or in other words
when the transitions do not carry a weight, Definition~\ref{def:WNWA}
is equivalent to the definition of a (unweighted) nested word
automaton \cite{AluMad06}. A language of nested words $L \subseteq
\NW(\Delta)$ is then called regular if it is accepted by a nested word
automaton. It is easy to see that this is the case iff the
characteristic series $\cha{L}: \NW(\Delta) \to \Bo$ is regular.

\begin{example}\label{ex:WNWA_appl}
  The procedure \texttt{bar} of Example~\ref{ex:nw_appl} can be
  modeled by a WNWA over $\K=([0,1],\max,\cdot,0,1)$ with four states
  $\{q_1,\ldots,q_4\}$. The transitions (only those with non-zero
  weight) are given as follows. We let $\iota(q_1)=1$ and
  $\kappa(q_4)=1$. Moreover,
  \begin{align*}
    &\deltaint(q_1,r,q_2)=1, \hspace{5em}\deltaint(q_2,b,q_3)
    =\deltaint(q_3,w,q_3)=\deltaint(q_3,ret,q_4)=1/2\\
    &\deltacall(q_2,call,q_1) =1/2, \hspace{5em}
    \deltaret(q_3,q_2,ret,q_3)=1/2. 
  \end{align*}
  Intuitively, each of the states corresponds to a line in the
  procedure \texttt{bar} which is the next to be executed. $q_1$
  corresponds to line $2$, $q_2$ corresponds to line $3$, $q_3$
  corresponds to line $7$ and $q_4$ is only reached at the end of an
  execution. Consider the nested word $nw$ of
  Example~\ref{ex:nw_appl}(2). There is exactly one run $r:q_1
  \stackrel{nw}{\to}q_4$ with $\weight(r) \ne 0$. We start in state
  $q_1$ execute $r$ and change to $q_2$. We then call and change back
  to $q_1$. After that we execute $r$ again and change to state $q_2$.
  We then execute $b$ and change to $q_3$. We return and stay in
  $q_3$.  Now we execute $w$ twice while staying in $q_3$ and finally
  end at state $q_4$. Observe that the automaton assigns $1/64$ to the
  nested word $nw$. 
\end{example}

\section{Weighted Logics}
\label{sec:mso}
In this section we introduce another formalism for specifying nested
word series.  For this we interpret a nested word $nw = (a_1 \ldots
a_n, \nu)$ as a relational structure consisting of the domain
$\dom(nw)=[n]$ together with the unary relations $\Lab_{a}=\{ i \in
\dom(nw)~|~ a_i = a\}$ for all $a \in \Delta$, the binary relation
$\nu$ and the usual $\leq$ relation on $\dom(nw)$.

First, we recall classical monadic second-order logic.  The set
$\MSO(\signw{\Delta})$ (we also write $\MSO$ for short) is given by
the following grammar.
$$
  \varphi\quad::=\quad x=y \mid \Lab_a(x) \mid x \leq y \mid \nu(x,y)
  \mid x\in X\mid \varphi\vee\phi\mid\neg\varphi \mid\exists
  x.\varphi\mid\exists X.\varphi
$$
where $a$ ranges over $\Delta$, where $x,y$ are first-order variables
and where $X$ is a second-order variable.  As usual we abbreviate $x <
y = \neg (y \leq x)$, $\phi \to \psi = \neg \phi \lor \psi$ and $\phi
\leftrightarrow \psi = {(\phi \to \psi)} \wedge {(\psi \to \phi)}$ for any
$\phi, \psi \in \MSO$.

Let $\phi \in \MSO$ and let $\Free(\phi)$ denote the set of variables
that occur free in $\phi$.  Let $\V$ be a finite set of first-order
and second-order variables such that $\Free(\phi) \subseteq \V$. A
\emph{$(\V,nw)$-assignment} $\gamma$ is a mapping from $\V$ to the
powerset $\PowS(\dom(nw))$ such that first-order variables are mapped
to singletons.  For $i \in \dom(nw)$ and $T \subseteq \dom(nw)$ we
denote by $\gamma[x \to i]$ (resp. $\gamma[X \to T]$) the $(\V\cup
\{x\}, nw)$-assignment (resp. $(\V\cup \{X\}, nw)$-assignment) which
equals $\gamma$ on $\V \setminus \{x\}$ (resp. $\V \setminus \{X\}$)
and assumes $\{i \}$ for $x$ (resp. $T$ for $X$).  We write
$(nw,\gamma) \models \phi$ if $\phi$ holds in $nw$ under the
assignment $\gamma$.  We write $\phi(x_1,\ldots ,x_n,X_1,\ldots,X_m)$
if $\Free(\phi) \subseteq \{ x_1,\ldots ,x_n,X_1,\ldots,X_m\}$.  In
this case write $nw \models \phi[i_1,\ldots ,i_n,T_1,\ldots,T_m]$
whenever we have $(nw, \gamma) \models \phi$ if $\gamma(x_j) =
\{i_j\}$ and $\gamma(X_j) = T_j$.  This is justified by the fact that
$(nw,\gamma) \models \phi$ only depends on the restriction
$\gamma_{|\Free(\phi)}$ of $\gamma$ to $\Free(\phi)$. Let
$\La_{\V}(\phi)=\{(nw,\gamma)~|~nw \in \NW(\Delta), \gamma \text{ is a
} (\V,nw)\text{-assignment, } (nw,\gamma)\models \phi\}$. Abbreviate
$\La(\phi)=\La_{\Free(\phi)}(\phi)$. Note that in case that $\phi$ is
a sentence, i.e. $\Free(\phi)=\emptyset$, we consider $\La(\phi)$ as a
subset of $\NW(\Delta)$.

Let $Z \subseteq \MSO$. A language $L \subseteq \NW(\Delta)$ is
\emph{Z-definable} if $L = \La(\phi)$ for a sentence $\phi \in Z$.
Formulae containing no quantification at all are called
\emph{propositional}. First-order formulae, i.e. formulae containing
only quantification over first-order variables are collected in $\FO$.
The class $\EMSO$ consists of all formulae $\phi$ of the form $\exists
X_1.  \ldots \exists X_m.  \psi$ where $\psi \in \FO$.  Alur and
Madhusudan showed that monadic second-order logic and nested word
automata are equally expressive.
\begin{theorem}[Alur \& Madhusudan~\cite{AluMad06,AluMad04b}]\label{thm:AM-regdefnm}
  A nested word language $L \subseteq \NW(\Delta)$ is regular iff $L$
  is $\MSO$-definable iff $L$ is $\EMSO$-definable.
\end{theorem}

We now turn to weighted monadic second-order logic as introduced
in~\cite{DroGas07}.  The set $\MSO(\K,\signw{\Delta})$ (once again we
shortly write $\MSO(\K)$) of \emph{weighted $\MSO$ formulae} over $\K$
is given by the following grammar:
\begin{align*}
  \phi \quad::=\quad &k\mid x=y\mid \Lab_a(x)\mid x \leq y\mid\nu(x,y)
  \mid x \in X \\* &\phantom{k} \mid \neg(x=y) \mid \neg
  \Lab_a(x)\mid \neg x \leq y\mid \neg\nu(x,y) \mid \neg (x \in X)\\*
  &\phantom{k}\mid \phi \lor \phi\mid \phi \wedge \phi\mid \exists
  x.\phi\mid \exists X.\phi\mid \forall x.\phi\mid \forall X.\phi
\end{align*}
where $k \in \K$, where $a$ ranges over $\Delta$, where $x,y$ are
first-order variables and where $X$ is a second-order variable.  Note
that we allow negation only for \emph{atomic formulae,} i.e. for the
formulae $x=y$, $\Lab_a(x)$, $x \leq y$, $\nu(x,y)$ and $x \in
X$. This is because in general semirings we do not have a natural
complement and hence it is not clear how to define the semantics of
negation for values other than $0$ and $1$ (cf. \cite{DroGas07}).

Let $\phi \in \MSO(\K)$ and $\Free(\phi) \subseteq \V$.  The weighted
semantics $\seman{\phi}_{\V}$ of $\phi$ is a function assigning a
value in $\K$ to a nested word $nw$ and a $(\V,nw)$-assignment
$\gamma$.  To each such pair $(nw,\gamma)$ we assign an element of
$\K$ inductively as follows. For $k \in \K$ we put
$\seman{k}_{\V}(nw,\gamma) = k$. For every other atomic formula or
negated atomic formula $\phi$ the semantics $\seman{\phi}_{\V}$ is
given by the characteristic function $\cha{\La_{\V}(\phi)}$.
Moreover, we define
\begin{align*}
  & \seman{\phi \lor \psi}_{\V} (nw,\gamma)& = & & &
  \seman{\phi}_{\V}(nw,
  \gamma) + \seman{\psi}_{\V}(nw,\gamma), \\
  & \seman{\phi \wedge \psi}_{\V} (nw,\gamma)& = & & &
  \seman{\phi}_{\V}(nw,
  \gamma) \cdot \seman{\psi}_{\V}(nw,\gamma), \\
  & \seman{\exists x. \phi}_{\V} (nw,\gamma)& = & & & \sum\nolimits_{i \in
    \dom(nw)}
  \seman{\phi}_{\V \cup \{x\} }(nw, \gamma[x \to i]), \\
  &\seman{\exists X. \phi}_{\V} (nw,\gamma)& = & & &\sum\nolimits_{T \subseteq
    \dom(nw)}
  \seman{\phi}_{\V \cup \{X\} }(nw, \gamma[X \to T]), \\
  &\seman{\forall x. \phi}_{\V} (nw,\gamma)& = & & &\prod\nolimits_{i \in
    \dom(nw)}
  \seman{\phi}_{\V \cup \{x\} }(nw, \gamma[x \to i]), \\
  &\seman{\forall X. \phi}_{\V} (nw,\gamma)& = & & &\prod\nolimits_{T \subseteq
    \dom(nw)} \seman{\phi}_{\V \cup \{X\} }(nw, \gamma[X \to T]). 
\end{align*}
We put $\seman{\phi} = \seman{\phi}_{\Free(\phi)}$.  Observe that in
the case where $\phi$ is a sentence, $\seman{\phi}$ can be considered
as a series from $\NW(\Delta)$ to $\K$.

\begin{remark}\label{rem:weight-boolean}
  A formula $\phi \in \MSO(\K)$ which does not contain a subformula $k
  \in \K$ can be interpreted as an unweighted formula.  We will use
  this implicitly in the sequel.  Moreover, note that if $\K$ is the
  Boolean semiring $\Bo$, then weighted logics and classical $\MSO$
  logic coincide.  In this case $k$ is either $0$ (false) or $1$
  (true).
\end{remark}

\begin{example}\label{ex:appl_wmso}\hfill
  \begin{enumerate}[(1)]
  \item As in Example~\ref{ex:nw_appl} suppose we model bibtex
    databases as nested words. Moreover, assume that $\text{tecrep}
    \in \Delta$ marks the beginning of an entry containing a technical
    report. Now, let $\K = \N$ be the semiring of the natural numbers.
    Then $(\seman{\exists x. \Lab_{\text{tecrep}}(x)},nw)$ counts the
    number of technical reports of the bibtex database modeled by
    $nw$.
  \item Again let $\K=\N$. Consider the formula $\phi=\forall
    x. \exists y.1$. Then $(\seman{\exists x.1},(a_1\ldots
    a_n,\nu))=n$ and $(\seman{\forall y. \exists x.  1},(a_1\ldots
    a_n,\nu))=n^n$. It can be shown as for words that $\seman{\phi}$
    is not regular as it grows too fast (cf. Example 3.4 in
    \cite{DroGas07}).
  \end{enumerate}
\end{example}



\noindent Let $Z \subseteq \MSO(\K)$. A series $S: \NW(\Delta) \to \K$
is \emph{Z-definable} if $S = \seman{\phi}$ for a sentence $\phi \in
Z$.  Example~\ref{ex:appl_wmso}(2) shows that unrestricted application
of universal quantification does not preserve regularity. Therefore we
now define different fragments of $\MSO(\K)$.

Note that the fragment $\RMSO(\K)$, the collection of
\emph{restricted} formulae, which was considered in~\cite{DroGas07}
and which characterizes regular formal power series is a semantic
restriction, and it is not clear whether membership in $\RMSO(\K)$ can
be decided.  In order to have a decidable fragment, we now
syntactically define the fragment $\sRMSO(\K)$. For this we follow the
approach of \cite{DroGas08}.

The idea is to restrict universal first-order quantification to
formulae having a semantics that takes on only finitely many values.
To this aim we start by identifying a class of formulae $\phi$ that
take on values $0$ and $1$ only, more precisely we will have
$\cha{\La_{\V}(\phi)}=\seman{\phi}_{\V}$.
The problem that arises is that by definition of the semantics, $\lor$
gets translated by means of $+$. Hence, for a formula $\phi = \phi_1
\lor \phi_2$ we only want to evaluate $\phi_2$ if $\phi_1$ evaluates
to $0$, otherwise we might end up with a sum greater than one.  A
similar problem occurs for $\exists x.$ and $\exists X.$

Given a classical (unweighted) $\MSO$-formula $\phi$ we assign to it
formulae $\phi^+$ and $\phi^-$ such that $\seman{\phi^+} =
\cha{\La(\phi)}$ and $\seman{\phi^-} = \cha{\La(\neg \phi)}$. The
crucial point is that we have a linear order at disposal.
\begin{enumerate}[(1)]
\item If $\phi$ is of the form $x=y$, $\Lab_a(x)$, $x \leq y$,
  $\nu(x,y)$, $x \in X$ then $\phi^+ = \phi$ and $\phi^- = \neg \phi$.
\item If $\phi = \neg \psi$, then $\phi^+ = \psi^-$ and $\phi^- =
  \psi^+$.
\item If $\phi = \psi \lor \psi'$, then $\phi^+ = \psi^+ \lor (\psi^-
  \wedge \psi'^+)$ and $\phi^-= \psi^- \wedge \psi'^-$.
\item If $\phi = \exists x. \psi(x)$, then $\phi^+ = \exists x.
  \psi(x)^+ \wedge \forall y. (y < x \wedge \psi(y))^-$ and $\phi^-=
  \forall x. \psi(x)^-$.
\end{enumerate}
In order to disambiguate set quantification, we have to define a
linear order on the subsets of the domain of a nested word or
equivalently on nested words (of fixed length) over the alphabet
$\{0,1\}$. We take the lexicographic order $<$ which is given by the
following formula. 
\begin{align*}
  X < Y = \exists y. y \in Y \wedge \neg y \in X \wedge \forall z. [z
  < y \to (z \in X \leftrightarrow z \in Y)]^+
\end{align*}
Now we proceed:
\begin{enumerate}[(1)]
\item[(5)] If $\phi = \exists X. \psi(X)$, then $\phi^+ = \exists X.
  \psi(X)^+ \wedge \forall Y.  (Y < X \wedge \psi(Y))^-$ and $\phi^-=
  \forall X. \psi(X)^-$.
\end{enumerate}

\noindent Formulae of the form $\phi^+$ or $\phi^-$ for some $\phi \in
\MSO$ are called \emph{syntactically unambiguous}.  Observe, if $\phi$
is syntactically unambiguous, then $\seman{\phi}_{\V} =
\cha{\La_{\V}(\phi)}$ for any finite set of variables $\V \supseteq
\Free(\phi)$. In the following, we shortly write $\phi \impl \psi$ for
$\phi^- \lor (\phi^+ \wedge \psi)$ for any two weighted formulae
$\phi, \psi$ where $\phi$ does not contain subformulae of the form $k$
($k \in \K$) and hence is also a classical formula.

We define $\aUMSO(\K)$, the collection of \emph{almost unambiguous}
formulae, to be the smallest subset of $\MSO(\K)$ containing all
constants $k$ ($k \in \K)$ and all syntactically unambiguous formulae
which is closed under conjunction and disjunction. Using the
distributivity, observe that for any $\psi \in \aUMSO(\K)$ there is a
formula $\psi'$ of the form $\psi'=\bigvee_{i=1}^n(k_i \wedge \psi_i)$
for some $k_i \in \K$ and syntactically unambiguous $\psi_i$ such that
$\seman{\psi} =\seman{\psi'}$ (cf.  \cite{DroGas08}).  We are now
ready to define the fragment $\sRMSO(\K)$.

\begin{definition}\label{def:srmso}
  A weighted formula $\phi$ is in $\sRMSO(\K)$ (syntactically
  restricted $\MSO$) if for every subformula $\theta$ of $\phi$ the
  following two conditions hold:
  \begin{enumerate}[(1)]
  \item If $\theta=\forall X. \psi$ for some $\psi \in \MSO(\K)$, then
    $\psi$ is syntactically unambiguous.
  \item If $\theta=\forall x. \psi$ for some $\psi \in \MSO(\K)$, then
    $\psi \in \aUMSO(\K)$.
  \end{enumerate}
\end{definition}

\noindent We collect in $\sRFO(\K)$ all $\phi\in \sRMSO(\K)$ which do
not contain any set quantification and we collect in $\sREMSO(\K)$ all
$\phi \in \sRMSO(\K)$ of the form $\exists X_1. \ldots \exists X_m.
\psi$ with $\psi \in \sRFO(\K)$.


Let now $\wUMSO(\K)$, the collection of \emph{weakly unambiguous}
formulae, be the smallest subset of $\MSO(\K)$ containing all
constants $k$ ($k \in \K)$ and all syntactically unambiguous formulae
which is closed under conjunction, disjunction and existential
quantification (both first- and second-order).  We define the fragment
$\swRMSO(\K)$.

\begin{definition}\label{def:swrmso}
  A weighted formula $\phi$ is in $\swRMSO(\K)$ (syntactically weakly
  restricted $\MSO$) if for every subformula $\theta$ of $\phi$ the
  following two conditions hold:
  \begin{enumerate}[(1)]
  \item If $\theta=\forall X. \psi$ for some $\psi \in \MSO(\K)$, then
    $\psi$ is syntactically unambiguous.
  \item If $\theta=\forall x. \psi$ for some $\psi \in \MSO(\K)$, then
    $\psi \in \wUMSO(\K)$.
  \end{enumerate}
\end{definition}
\noindent Clearly, $\aUMSO(\K) \subset \wUMSO(\K) \subset \sRMSO(\K)
\subset \swRMSO(\K) \subset \MSO(\K)$.  The first main result of this
paper is the characterization of regular nested word series using
weighted logics.  It reads as follows.

\begin{samepage}
\begin{theorem}\label{thm:nw-main}
  Let $\K$ be a commutative semiring and let $S:\NW(\Delta) \to \K$ be
  a nested word series. Then the following holds.
  \begin{enumerate}[\em(a)]
  \item $S$ is regular iff it is $\sRMSO(\K)$-definable iff it is
    $\sREMSO(\K)$-definable.
  \item If $\K$ is additively locally finite, then $S$ is regular iff
    it is $\swRMSO(\K)$-definable.
  \item If $\K$ is locally finite, then $S$ is regular iff it is
    $\MSO(\K)$-definable.
  \end{enumerate}
\end{theorem}
\end{samepage}

\noindent We prove the result at the end of
Section~\ref{sec:interpr-nest-words} by interpreting nested words in
alternating texts. In the next section we introduce alternating texts
and weighted automata for them.

\begin{example}\label{ex:nestdepth}
  The \emph{nesting depth} of a position $i$ of a nested word $nw$ is
  the number of open call positions (i.e. where the corresponding
  return position has not occurred yet including the position itself).
  The nesting depth of a nested word is the maximum nesting depth of
  its positions.  Let $\K = (\Z \cup \{-\infty\},\max,+, -\infty,0)$.
  \begin{align*}
    &\text{open}(x) = \forall y. (y \leq x \wedge \call(y)) \impl 1
    \land (y \leq x \wedge \return(y)) \impl -1 ~~~\text{where} \\*
    &\call(x) = \exists y. \nu(x,y) ~~~\text{and } \return(x) =
    \exists y. \nu(y,x)
  \end{align*}
  Then $\seman{\exists x. \text{open}(x)}$ assigns to a nested word
  its nesting depth. Hence, since $\exists x. \text{open}(x)\in
  \sRMSO(\K)$, the series is regular by Theorem~\ref{thm:nw-main}.
\end{example}
  
\section{Alternating Texts}
\label{sec:spb}

A bisemigroup is a set together with two associative operations.
Several authors investigated the free bisemigroup as a fundamental,
two-dimensional extension of classical automaton theory, see e.g.
{\'E}sik and N{\'e}meth \cite{EsiNem04} and Hashiguchi et al.  (e.g.
\cite{Hasetal00,Hasetal02,Hasetal03}).  {\'E}sik and N{\'e}meth
considered as a representation for the free bisemigroup the so-called
\emph{sp-biposets}, a certain class of biposets.  A different
representation of the free bisemigroup over some finite set $\Delta$
are the so-called alternating texts~\cite{EhrRoz93,HooPas97}.  A
\emph{text} over $\Delta$ is a tuple $(V, \lambda, \leq_1, \leq_2)$
where $\leq_1$ and $\leq_2$ are linear orders over a finite but
non-empty domain $V$ and $\lambda: V \to \Delta$ is a labeling
function. Of course we consider texts only up to isomorphism.
Therefore, unless otherwise specified, the domain of a text will be
$[n]=\{1,\ldots,n\}$ for some $n \in\N$ and $\leq_1$ will correspond
to the canonical order on $[n]$.

We define the binary operations $\circ$ and $\bullet$, called the
horizontal and vertical product, on texts as follows: Let $\tau=(V,
\lambda, \leq_1, \leq_2)$ and $\tau'=(V', \lambda', \leq'_1, \leq'_2)$
be two texts where we assume that $V$ and $V'$ are disjoint. Then
\begin{align*}
  \tau_1 \circ \tau_2 &= (V \disjoint V',\lambda \cup \lambda', \leq_1
  \cup \leq'_1\cup V \times V', \leq_2 \cup \leq'_2 \cup V
  \times V'), \\
  \tau_1 \bullet \tau_2 &= (V \disjoint V', \lambda \cup \lambda',
  \leq_1 \cup\leq'_1\cup V \times V', \leq_2 \cup \leq'_2 \cup V'
  \times V).
\end{align*}
\bigskip

  \begin{center}
    \begin{picture}(112,22)(0,-3) \nullfont
      \node[Nh=4,Nw=4](n0)(0,10){$a$} 
      \node[Nh=4,Nw=4](n1)(16,10){$a$}
      \node[Nw=4,Nh=4](n2)(32,10){$c$}
      \node[Nw=4,Nh=4](n3)(48,10){$a$}
      \node[Nh=4,Nw=4](n4)(64,10){$c$}
      \node[Nh=4,Nw=4](n5)(80,10){$a$}
      \node[Nw=4,Nh=4](n6)(96,10){$b$}
      \node[Nw=4,Nh=4](n7)(112,10){$b$}
      \node[Nh=0,Nw=0,Nframe=n](h0)(0,18){}
      \node[Nh=0,Nw=0,Nframe=n](h1)(16,18){}
      \node[Nw=0,Nframe=n,Nh=0](h2)(32,18){}
      \node[Nw=0,Nframe=n,Nh=0](h3)(48,18){}
      \node[Nh=0,Nw=0,Nframe=n](h4)(64,15){}
      \node[Nh=0,Nw=0,Nframe=n](h5)(80,15){}
      \node[Nw=0,Nframe=n,Nh=0](h6)(96,15){}
      \node[Nw=0,Nframe=n,Nh=0](h7)(112,18){}
    
      \drawedge[linewidth=0.17](n1,n0){}
      \drawedge[linewidth=0.17](n3,n2){}
      \drawedge[linewidth=0.17](n4,n5){}
      \drawedge[linewidth=0.17](n5,n6){}

      \drawedge[linewidth=0.17,curvedepth=-12](n0,n7){}
      \drawedge[linewidth=0.17,curvedepth=8](n6,n3){}
      \drawedge[linewidth=0.17,curvedepth=-8](n7,n4){}    
  \end{picture}
\end{center}

\begin{enumerate}[Figure 2:]
\item
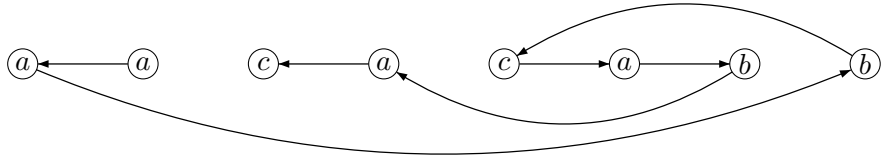
  A visualization of the alternating text given by $(a\bullet
  a)\circ (c\bullet a \bullet (c \circ a \circ b)\bullet b)$. Here we
  only give the successor relation of the second order. The first
  order is given simply from the left to the right.
  \label{fig:text}
\end{enumerate}

Let $\TXT(\Delta)$ be the class of texts which can be obtained from
the singleton texts by finite applications of $\circ,\bullet$. This
class was named the class of \emph{alternating texts} in
\cite{EhrRoz93}.  The class $\TXT(\Delta)$ together with the
operations $\circ, \bullet$ is the free bisemigroup over
$\Delta$~\cite{HooPas97}. Let monadic second-order logic
$\MSO(\sigtxt{\Delta})$ and weighted logics for texts, denoted
\mbox{$\MSO(\K,\sigtxt{\Delta})$} be defined along the same lines as
for nested words.  Moreover, define $\sRMSO(\K,\sigtxt{\Delta})$ and
$\swRMSO(\K,\sigtxt{\Delta})$ using the linear order $\leq_1$.

Now we introduce weighted parenthesizing automata (cf. \cite{Mat07})
operating on the free bisemigroup generalizing parenthesizing automata
as introduced by {\'E}sik and N{\'e}meth \cite{EsiNem04}.

\begin{definition}\label{def:wbpa}
  A tuple $\A=(\HS,\VS,\Omega, \mu,\muop,\mucl,\lambda, \gamma)$ is a 
  \emph{weighted parenthesizing automaton}~(WPA) provided that
  \begin{enumerate}[$\bullet$]
  \item $\HS$ and $\VS$ are finite, disjoint sets of \emph{horizontal}
    and \emph{vertical states}, respectively,
  \item$\Omega$ is a finite set of \emph{parentheses},
    \footnote{
      We let $s \in \Omega$ represent both an opening and a closing
      parentheses. To help the intuition we also write~$(_s$~or~$)_s$
      for $s$.
    }
  \item $\mu: (\HS \times \Delta \times \HS) \cup (\V \times \Delta
    \times \V) \to \K$ is the \emph{transition function},
  \item $\muop, \mucl: (\HS \times \Omega \times \V) \cup (\V \times
    \Omega \times \HS) \to \K$ are the \emph{opening} and
    \emph{closing parenthesizing functions}, respectively,
  \item $\lambda, \gamma: \HS \cup \V \to \K$ are the \emph{initial}
    and \emph{final weight functions}, respectively.
\end{enumerate}
\end{definition}

\noindent We now come to the notion of a run $r$ of $\A$.  We given an
inductive definition where we also define its \emph{label} $\lab(r)
\in \TXT(\Delta)$, its \emph{weight} $\weight_{\A}(r)\in \K$, its
\emph{initial state} $\initial(r) \in \HS \cup \VS$ and its
\emph{final state} $\final(r) \in \HS \cup \VS$. Formally the set of
\emph{runs} of $\A$ is the smallest set of words over the alphabet
$\Delta \cup \Omega \cup \HS \cup \VS \cup\{(,)\}\cup \{,\}$ such
that:
\begin{enumerate}[(1)]
\item The word $(q_1,a,q_2)$ is a run for all $(q_1,q_2) \in (\HS
  \times \HS) \cup (\V \times \V)$ and $a \in \Delta$. We set
  \begin{align*} 
    &\lab((q_1,a,q_2))=a \in \TXT(\Delta),
    \hspace{2em} \weight_{\A}((q_1,a,q_2))= \mu(q_1,a,q_2),\\
    & \initial((q_1,a,q_2)) = q_1 \text{ and }
    \final((q_1,a,q_2))=q_2.&
  \end{align*}
\item If $r_1$ and $r_2$ are runs such that $\final(r_1) =
  \initial(r_2) \in \HS$ (respectively such that $\final(r_1) =
  \initial(r_2) \in \V$), then $r = r_1 r_2$ is a run having
  \begin{align*}
    &\lab(r) =\lab(r_1) \circ \lab(r_2), & (\text{resp. }
    \lab(r)
    =\lab(r_1) \bullet \lab(r_2)),\\
    &\weight_{\A}(r) = \weight_{\A}(r_1) \cdot \weight_{\A}(r_2),
    &\initial(r) = \initial(r_1) \text{ and } \final(r)=\final(r_2). 
  \end{align*}
\item If a run $r$ resulting from 2 has $\initial(r) \in \HS$
  (resp. $\initial(r) \in \V$) and if $q_1,q_2 \in \VS$ (resp. if
  $q_1,q_2 \in \HS$) and $s \in \Omega$, then $r' = (q_1, (_s,
  \initial(r)) ~r~ (\final(r), )_s, q_2)$ is a run. We set
  \begin{align*}
    & \lab(r') =\lab(r), \hspace{2em} \initial(r') = q_1 \text{ and } \final(r')=q_2,\\*
    & \weight_{\A}(r') = \muop((q_1, (_s, \initial(r))) \cdot
    \weight_{\A}(r) \cdot \mucl((\final(r), )_s, q_2)). &
  \end{align*}
\end{enumerate}

\noindent Let $\tau \in \TXT(\Delta)$. Since in (3) above we require
that the run $r$ we start with results from (2), we do not allow
repeated application of (3) and therefore there are only finitely many
runs $r$ of $\A$ with label $\tau$.  Intuitively, we do not allow for
doubled parentheses.  If $r$ is a run of $\A$ with $\lab(r) =\tau$,
$\initial(r)=q_1$, $\final(r)=q_2$, we write $r: q_1
\stackrel{\tau}{\to} q_2$.  The behavior of $\A$ is a text series
$\behave{\A}: \TXT(\Delta) \to \K$. It is given by
\begin{align*}
  (\behave{\A},\tau) = \sum_{q_1,q_2 \in \HS \cup \V} \lambda(q_1)
  \cdot \sum_{r: q_1 \stackrel{\tau}{\to} q_2} \weight_{\A}(r) \cdot
  \gamma(q_2).
\end{align*}
A text series $S$ is \emph{regular} if there is a WPA $\A$ such that
$\behave{\A} = S$.

\begin{theorem}[see \cite{Mat09b}]
\label{thm:txt-main}\label{thm:txt-srmso}\label{txt-swrmso}
Let $\K$ be a commutative semiring and let $S:\TXT(\Delta) \to \K$ be
an alternating text series. Then the following holds.
\begin{enumerate}[(a)]
\item $S$ is regular iff it is $\sRMSO(\K)$-definable iff it is
  $\sREMSO(\K)$-definable.
\item If $\K$ is additively locally finite, then $S$ is regular iff it
  is $\swRMSO(\K)$-definable.
\item If $\K$ is locally finite, then $S$ is regular iff it is
  $\MSO(\K)$-definable.
\end{enumerate}
\end{theorem}

\noindent We note that the proof in \cite{Mat09b} is effective, i.e.
given an $\sRMSO(\K)$ (resp. $\swRMSO(\K)$, resp.  $\MSO(\K)$) formula
$\phi$ we can effectively construct a WPA $\A$ such that $\seman{\phi}
= \behave{\A}$, and conversely, given a WPA $\A$ we can effectively
construct $\phi \in \sREMSO(\K)$ such that $\seman{\phi} =
\behave{\A}$.

\section{Interpreting Nested Words in Alternating Texts}
\label{sec:interpr-nest-words}

We will now derive similar results for nested words as for alternating
texts by interpreting the different structures within each other. For
this we utilize definable transductions as introduced by
Courcelle~\cite{Cou94}. We only have to ensure that they preserve
definability, now with respect to weighted logics.  First, we
introduce the notion of definable transductions. For this let
$\sigma_1$ and $\sigma_2 = ((R_i)_{i \in I}, \rho)$ be two relational
signatures where $\rho: I \to \N_+$ assigns to each relation symbol
$R_i$ a positive arity. Moreover, let $\C_1$ and $\C_2$ be classes of
finite $\sigma_1$- and $\sigma_2$-structures, respectively.  Let
monadic second-order logic $\MSO(\sigma_1)$ and $\MSO(\sigma_2)$ be
defined along the lines as for nested words.

By a \emph{$(\sigma_1,\sigma_2)$-$1$-copying definition scheme with
  parameters $X_1, \ldots, X_n$} we mean a tuple $\D = (\theta,\delta,
(\phi_i)_{i \in I})$ of formulae in $\MSO(\sigma_1)$ such that
$\Free(\theta)\subseteq\{X_1,\ldots,X_n\}$, $\Free(\delta) \subseteq
\{x_1,X_1,\ldots,X_n\}$ and $\Free(\phi_i) \subseteq\{x_1, \ldots ,
x_{\rho(i)},X_1,\ldots X_n\}$ for all $i \in
I$. 

Let $\D$ be a $(\sigma_1,\sigma_2)$-$1$-copying definition scheme, let
$s_1 \in \C_1$ and let $T_1,\ldots,T_n$ subsets of the domain
$\dom(s_1)$ of $s_1$ such that $s_1 \models
\theta[T_1,\ldots,T_n]$. Then define the $\sigma$-structure
$\deftrans_{\D}(s_1,T_1,\ldots,T_n)=s_2$ with domain
$\dom(s_2)\subseteq \dom(s_1)$ and interpretations of relation symbols
$R_i^{s_2}$ given as follows:
\begin{align*}
  v \in \dom(s_2) \Leftrightarrow &~ s_1 \models
  \delta[v,T_1,\ldots,T_n]
  \text{ for all } v \in \dom(s_1).  \\*
  (v_1, \ldots , v_{\rho(i)}) \in R_i^{s_2} \Leftrightarrow &~ s_1
  \models \phi_{i}[v_1, \ldots, v_{\rho(i)},T_1,\ldots,T_n] \text{ for
    all } i \in I\text{ and }\\& \hspace{14.5em} \text{all } v_1,
  \ldots, v_{\rho(i)} \in \dom(s_2).
\end{align*}
By abusing notation, we define the transduction
$\deftrans_{\D}\subseteq \C_1 \times \C_2$ by letting $(s_1,s_2) \in
\deftrans_{\D}$ iff $s_1 \in \C_1$ and there are sets $T_1,\ldots,
T_n\subseteq \dom(s_1)$ with $s_1\models \theta[T_1,\ldots,T_n]$ such
that $s_2=\deftrans_{\D}(s_1)$. Let us call a definition scheme $\D$
with parameters $X_1,\ldots,X_n$ \emph{unambiguous} if for any pair
$(s_1,s_2)\in \deftrans_{\D}$ there is at most one assignment of
parameters $\gamma:\{X_1,\ldots,X_n\} \to \PowS(\dom(s_1))$ such that
$\deftrans_{\D}(s_1,\gamma(X_1),\ldots,\gamma(X_n))=s_2$.

\begin{definition}\label{def:deftrans}  
  A transduction $\Phi\subseteq \C_1 \times \C_2$ is
  \emph{unambiguously definable} if there is a unambiguous definition
  scheme $\D$ such that $\Phi = \deftrans_{\D}$. It is unambiguously
  $\FO$-definable if there is an unambiguous definition scheme $\D =
  (\theta,\delta, (\phi_i)_{i \in I})$ defining $\Phi$ with
  $\theta,\delta, (\phi_i)_{i \in I} \in \FO$.
\end{definition}

A transduction which is given by a less restricted definition scheme,
where one allows for more than one copy of $s_1$ and which is not
necessarily unambiguous, is called definable.  Courcelle~\cite{Cou94}
showed that the preimage of a definable set under a definable
transduction is again definable. We will show a similar result for
series.  Let $\Phi: \C_1 \to \C_2$ be a partial function with domain
$\dom(\Phi)$ and let $S: \C_2 \to \K$. Define $\Phi^{-1}(S)$ by
letting $(\Phi^{-1}(S),s_1) = (S, \Phi(s_1))$ for all $s_1 \in
\dom(\Phi)$ and $(\Phi^{-1}(S),s_1) = 0$ otherwise.  If $\Phi$ is
injective, we let $\Phi(S)=(\Phi^{-1})^{-1}(S)$.

Clearly, $\MSO(\K)$ can be defined for $\C_1$ and $\C_2$ along the
same lines as for nested words. In order to disambiguate a formula, we
need a linear order on each $s \in \C_1$ (resp. $\C_2$).  For the next
proposition we therefore assume that there are binary relation symbols
$\leq_1 \in \sigma_1$ and $\leq_2 \in \sigma_2$ such that the
interpretation of $\leq_i$ in $s$ is a linear order for any $s \in
\C_i$ $(i=1,2)$. Using these linear orders we can define syntactically
unambiguous formulae and then $\sRMSO(\K)$ and $\swRMSO(\K)$ over
$\sigma_1$ and $\sigma_2$.
\begin{proposition}\label{prop:deftrans-weightdef}
  Let $\Phi: \C_1 \to \C_2$ be an unambiguously definable partial
  function.  Then the following holds:
  \begin{enumerate}[\em(1)]
  \item If $S: \C_2 \to \K$ is $\MSO(\K)$-definable, then so is
    $\Phi^{-1}(S)$.
  \item If $S: \C_2 \to \K$ is $\sRMSO(\K)$-definable, then so is
    $\Phi^{-1}(S)$.
  \item If $S: \C_2 \to \K$ is $\swRMSO(\K)$-definable, then so is
    $\Phi^{-1}(S)$.
  \item If $\Phi$ is unambiguously $\FO$-definable and $S: \C_2 \to
    \K$ is $\sREMSO(\K)$-definable, then $\Phi^{-1}(S)$ is
    $\sREMSO(\K)$-definable.
\end{enumerate}
\end{proposition}
\begin{proof}[Proof sketch. Full proof and more general results can be
  found in~\cite{Mat09b,Mat09a}]\hfill

  \noindent Let $\D = (\theta, \delta, (\phi_i)_{i \in I})$ be an unambiguous
  definition scheme defining $\Phi$.  Let $\phi \in \MSO(\K)$.  By
  induction on the structure of $\phi$ we now define the formula
  $\widehat{\phi} \in \MSO(\K,\sigma_1)$.
  \begin{align*}
    \widehat{k} & = k, & \widehat{x = y} & = (x = y) & \widehat{x \in
      X} & = x \in X & \widehat{R_i(x_1, \ldots, x_{\rho(i)})} & =
    \phi_{i}(x_1 \ldots x_{\rho(i)},X_1,\ldots,X_n)^+
  \end{align*}
  If $\phi$ is $x=y$, $x \in X$ or $R_i(x_1, \ldots,x_{\rho(i)})$ let
  $\widehat{\neg \psi} = (\widehat{\psi})^-$. Moreover, let
  \begin{align*}
    \widehat{\psi_1 \wedge \psi_2} &=
    \widehat{\psi_1} \wedge \widehat{\psi_2} \\
    \widehat{\psi_1 \vee \psi_2} &=
    \begin{cases}
      (\widehat{\psi_1} \vee \widehat{\psi_2})^+ &\text{if } \psi_1
      \lor \psi_2
      \text{ is syntactically unambiguous} \\
      \widehat{\psi_1} \vee \widehat{\psi_2} &\text{otherwise}
    \end{cases}
    \\
    \widehat{\exists x. \psi} &=
    \begin{cases}
      [\exists x. (\delta(x,X_1,\ldots,X_n) \wedge
      \widehat{\psi})]^+&\text{if }
      \exists x. \psi \text{ is syntactically unambiguous} \\
      \exists x. (\delta(x,X_1,\ldots,X_n)^+ \wedge \widehat{\psi})
      &\text{otherwise}
    \end{cases}
    \\
    \widehat{\exists X. \psi} &=
    \begin{cases}
      [\exists X. \forall x.  (x \in X \to \delta(x,X_1,\ldots,X_n))
      \wedge \widehat{\psi}]^+&\text{if }
      \exists X. \psi \text{ is synt. unambiguous} \\
      \exists X. \forall x.  (x \in X \to \delta(x,X_1,\ldots,X_n))^+
      \wedge \widehat{\psi} &\text{otherwise}
    \end{cases}
    \\
    \widehat{\forall x. \psi} &= \forall x.  \delta(x,X_1,\ldots,X_n)
    \impl \widehat{\psi}
    \\
    \widehat{\forall X. \psi} &= \forall X.\left(\forall x.  x \in X
      \to \delta(x,X_1,\ldots,X_n) \right) \impl \widehat{\psi}.
\end{align*}
Now let $\phi$ be as required such that $\seman{\phi} = S$.  One can
show by induction on the structure of $\phi$ that $\seman{\exists
  X_1,\ldots X_n. \theta(X_1,\ldots,X_n)^+ \wedge \widehat{\phi}} =
\Phi^{-1}(S)$.  By construction we get that if $\phi$ is syntactically
unambiguous, then so is its translation $\widehat{\phi}$. Again by
induction it is therefore not hard to see that $\widehat{\phi}$ is in
$\aUMSO(\K)$ (resp.  $\wUMSO(\K)$) if $\phi$ is in $\aUMSO(\K)$
(resp. $\wUMSO(\K)$).  From this we conclude that the translation is
as required.\eop
\end{proof}

We are now going to show that regular series coincide with
$\sRMSO(\K)$-definable ones. For this we define two embeddings of
nested words into alternating texts and use the characterizations of
text series. The connection we establish turns out to be useful again
in Section~\ref{cha:algebraic}.  Define $\Phi_\bullet, \Phi_\circ:
\NW(\Delta) \to \TXT(\Delta)$ as follows.  Let $nw = ( w, \nu) \in
\NW(\Delta)$ where $w=a_1 \ldots a_n$. If $\nu = \emptyset$, then let
$\Phi_\circ(nw) = a_1 \circ \ldots \circ a_n$ and $\Phi_\bullet(nw) =
a_1 \bullet \ldots \bullet a_n$. If $\nu \ne \emptyset$, let $i$ be
the minimal call position and $j$ the corresponding return position.
Let $nw'=nw[i+1,j-1]$ and $nw''=nw[j+1,n]$. Suppose for the moment
that $i+1 \leq j-1$ and $j+1 \leq n$. We define
\begin{align*}
  \Phi_\circ(nw) &=a_1 \circ \ldots \circ a_{i-1} \circ (
  a_i\bullet \Phi_\bullet(nw') \bullet a_j) \circ \Phi_\circ(nw''), \\
  \Phi_\bullet(nw) &=a_1 \bullet \ldots \bullet a_{i-1} \bullet ( a_i
  \circ \Phi_\circ(nw') \circ a_j) \bullet \Phi_\bullet(nw'').
\end{align*}
If $i+1=j$ or $j=n$, then we just ignore the terms $\Phi_\circ(nw')$,
$\Phi_\circ(nw'')$, $\Phi_\bullet(nw')$ and $\Phi_\bullet(nw'')$,
respectively, in the definition above.  Intuitively, we transform the
nesting relation into well-matched brackets. As an example consider
the nested word $nw$ given in Figure~\ref{fig:nw}. Its coding
$\Phi_\circ(nw)$ is the alternating text in Figure~2.

Let $\Phi_\circ(nw) =
(V^\circ,\lambda^\circ,\leq^\circ_1,\leq^\circ_2)$ and
$\Phi_\bullet(nw) = (V^\bullet,\lambda^\bullet,
\leq^\bullet_1,\leq^\bullet_2)$. The following observations can easily
be made by induction either on $n$ or on $|\nu|$:
\begin{enumerate}[(a)]
\item Both $V^\circ$ and $V^\bullet$ have cardinality $n$. We
  therefore assume from now on that $V^\circ=V^\bullet=[n]$ such that
  $\leq^\circ_1$ as well as $\leq^\bullet_1$ is the usual order on
  $[n]$. It is easy to see that
  $\lambda^\circ(i)=\lambda^\bullet(i)=a_i$.\label{enum:phi-observ1}
\item Both $\Phi_\circ$ and $\Phi_\bullet$ are
  injective. \label{enum:phi-observ2}
\end{enumerate}

\noindent Recall that a position of $nw$ has odd \emph{nesting depth}
if the number of open call positions is odd (see
Example~\ref{ex:nestdepth}).

\begin{lemma}\label{lem:trans-evennestdep}
  Let $nw = (a_1 \ldots a_n, \nu)\in \NW(\Delta)$, let $\Phi_\circ(nw)
  = ([n],\lambda, \leq^\circ_1,\leq^\circ_2)$ and let $
  \Phi_\bullet(nw) = ([n],\lambda,\leq^{\bullet}_1,\leq^{\bullet}_2)$.
  Moreover, let $1 \leq i< j\leq n$.  Then we have, $i \geq^\circ_2 j$
  iff $i\leq^{\bullet}_2 j$ iff there is some $(k,\ell) \in \nu$ with $1
  \leq k \leq i < j \leq \ell \leq n$ such that there is no $(k',\ell') \in
  \nu$ with $k <k'\leq i< j\leq \ell'<\ell$ and $k$ has odd nesting depth.
\end{lemma}
\begin{proof}
  The proof is by induction on $|\nu|$. For $|\nu|=0$ this is trivial.
  Now let $|\nu|\geq 1$.  We only prove that $i \geq^\circ_2 j$ iff
  there is some $(k,\ell) \in \nu$ with $1 \leq k \leq i < j \leq \ell \leq
  n$ such that there is no $(k',\ell') \in \nu$ with $k <k'\leq i< j\leq
  \ell'<\ell$ and $k$ has odd nesting depth.  That this holds iff
  $i\leq^{\bullet}_2 j$ can be shown analogously.  Let $i'$ be the
  minimal call position and $j'$ the corresponding return position.
  Let $nw'=nw[i'+1,j'-1]$ and $nw''=nw[j'+1,n]$ provided they exist.
  Moreover, let $\Phi_\bullet(nw') = ([j'-i'-1],\lambda',
  \leq'_1,\leq'_2)$ and $\Phi_\circ(nw'') = ([n-j'],\lambda'',
  \leq_1'',\leq_2'')$.  We consider three cases:
  \begin{enumerate}[(1)]
  \item Assume $i<i'$ or $i \leq j'< j$.  Then $i \leq^\circ_2 j$
    and there is no $(k,\ell) \in \nu$ with $1 \leq k \leq i < j \leq
    l \leq n$. 
  \item Assume $i'\leq i < j \leq j'$. If $i=i'$ or $j=j'$, then $i
    \geq^\circ_2 j$ and choosing $(k,\ell)=(i,j)$ gives $(k,\ell)$ as
    required since $i$ has nesting depth 1. If $i'<i < j< j'$, then we
    get:
    \begin{align*}
      i\geq^\circ_2 j &\hspace{1em}\Longleftrightarrow\hspace{1em} i-i' \geq'_2 j-i'\\
      &\hspace{1em}\Longleftrightarrow\hspace{1em} \text{not } i-i' \leq'_2 j-i'\\[0.2cm]
      &\hspace{1em}\Longleftrightarrow\hspace{1em}
      \parbox{9.5cm}{either there is some $(k,\ell) \in \nu[i'+1,j'-1]$
        with $1 \leq k \leq i-i' < j-i' \leq \ell \leq j'-i'-1$ such that
        there is no $(k',\ell') \in \nu[i'+1,j'-1]$ with $k <k'\leq i-i'
        < j-i'\leq \ell'<\ell$ and $k$ has even nesting depth in $nw'$, or
        there is no $(k,\ell) \in \nu[i'+1,j'-1]$ with $1 \leq k \leq
        i-i' < j-i'
        \leq \ell \leq j'-i'-1$}\\[0.2cm]
      &\hspace{1em}\Longleftrightarrow\hspace{1em}
      \parbox{9.5cm}{there is some $(k,\ell) \in \nu$ with $1 \leq k
        \leq i < j \leq \ell \leq n$ such that there is no $(k',\ell') \in
        \nu$ with $k <k'\leq i< j\leq \ell'<\ell$ and $k$ has odd nesting
        depth.}
    \end{align*}
  \item Assume $j'<i$. Then we get
    \begin{align*}
      i\geq^\circ_2 j &\hspace{1em}\Longleftrightarrow\hspace{1em}i-j' \geq''_2
      j-j'\\[0.2cm] &\hspace{1em}\Longleftrightarrow\hspace{1em}
      \parbox{9.5cm}{there is some $(k,\ell) \in \nu[j'+1,n]$ with $1
        \leq k \leq i-j' < j-j' \leq \ell \leq n-j'$ such that there is
        no $(k',\ell') \in \nu[j'+1,n]$ with $k <k'\leq
        i-j'< j-j'\leq \ell'<\ell$ and $k$ has odd nesting depth}\\[0.2cm]
      &\hspace{1em}\Longleftrightarrow\hspace{1em}
      \parbox{9.5cm}{there is some $(k,\ell) \in \nu$ with $1 \leq k
        \leq i < j \leq \ell \leq n$ such that there is no $(k',\ell') \in
        \nu$ with $k <k'\leq i<
        j\leq \ell'<\ell$ and $k$ has odd nesting depth.}\\[-4.5ex]
    \end{align*} 
  \end{enumerate}\eop
\end{proof}

\begin{corollary}\label{cor:nwspb-phi-def}
  The functions $\Phi_\circ$ and $\Phi_\bullet$ are unambiguously
  $\FO$-definable.
\end{corollary}
\begin{proof}
  We only show that $\Phi_\circ$ is $\FO$-definable. For
  $\Phi_\bullet$ the claim can be shown analogously.  We give a
  1-copying definition scheme $(\theta,\delta, (\phi_{\Lab_a})_{a\in
    \Delta}, \phi_{\leq_1},\phi_{\leq_2})$ with four parameters
  $X_1,X_2,Y_1,Y_2$.

  Let the macros $\call(x)$ and $\return(x)$ be as in
  Example~\ref{ex:nestdepth}. Moreover, let
  \begin{align*}
    \frst_{\nu}(x) = & \call(x)\wedge\forall y.\call(y)\to x\leq y
  \end{align*}
  The next macro defines $y$, the next call or return position
  following position $x$.
  \begin{align*}
    \next_{\nu}(x,y) = &x < y \wedge (\call(y) \lor \return(y)) \wedge
    \forall z. (x < z < y) \to (\neg \call(z) \land \neg\return(z))
  \end{align*}
  We now define the formula $\theta(X_1,X_2,Y_1,Y_2)$ which for all
  $nw=(a_1\ldots a_n,\nu) \in \NW(\Delta)$ and $C_1,C_2,R_1,R_2
  \subseteq [n]$ has the property that $nw
  \models\theta[C_1,C_2,R_1,R_2]$ iff $C_1$ is the set of all call
  positions of odd nesting depth, $C_2$ is the set of all call
  positions of even nesting depth, $R_1$ is the set of all return
  positions of even nesting depth and $R_2$ is the set of all return
  positions of odd nesting depth.
  \begin{align*}
    \theta(X_1,X_2,Y_1,Y_2) = &~ (X_1 \cap X_2 = \emptyset) \wedge
    \forall
    z.  (z \in X_1 \lor z \in X_2) \to \call(z) ~ \\
    &\wedge (Y_1 \cap Y_2 = \emptyset) \wedge \forall z. (z
    \in Y_1 \lor z \in Y_2) \to \return(z)  \\
    &\wedge~ \forall z.
    \frst_{\nu}(z) \to z \in X_1\\
    &\wedge \forall z_1,z_2.( (z_1 \in X_1 \wedge \next_{\nu}(z_1,z_2)
    \wedge \return(z_2)) \to z_2
    \in Y_1)\\
    &\wedge \forall z_1,z_2.( (z_1 \in X_1 \wedge \next_{\nu}(z_1,z_2)
    \wedge \call(z_2)) \to z_2
    \in X_2) \\
    &\wedge \forall z_1,z_2.( (z_1 \in X_2 \wedge \next_{\nu}(z_1,z_2)
    \wedge \return(z_2)) \to z_2
    \in Y_2)\\
    &\wedge \forall z_1,z_2.( (z_1 \in X_2 \wedge \next_{\nu}(z_1,z_2)
    \wedge
    \call(z_2)) \to z_2 \in X_1) \\
    &\wedge \forall z_1,z_2.( (z_1 \in Y_1 \wedge \next_{\nu}(z_1,z_2)
    \wedge \return(z_2)) \to z_2
    \in Y_2) \\
    &\wedge \forall z_1,z_2.( (z_1 \in Y_1 \wedge \next_{\nu}(z_1,z_2)
    \wedge \call(z_2)) \to z_2
    \in X_1) \\
    &\wedge \forall z_1,z_2.( (z_1 \in Y_2 \wedge \next_{\nu}(z_1,z_2)
    \wedge \return(z_2)) \to z_2
    \in Y_1) \\
    &\wedge \forall z_1,z_2.( (z_1 \in Y_2 \wedge \next_{\nu}(z_1,z_2)
    \wedge \call(z_2)) \to z_2 \in X_2)
  \end{align*}
  where $X \cap Y=\emptyset$ abbreviates $\neg(\exists z. z \in X
  \land z \in Y)$. We let $\delta(x,X_1,X_2,Y_1,Y_2)$ be some
  tautology.  Now we define the interpreting formulae.  We set
  $\phi_{\Lab_{a}}(x,X_1,X_2,Y_1,Y_2)) = \Lab_a(x)$ and let
  $\phi_{\leq_1}(x,y,X_1,X_2,Y_1,Y_2))=x\leq y$. Furthermore, we
  define $\phi_{\circ}(x,y,X_1)$ to be the following formula which
  expresses the condition of Lemma~\ref{lem:trans-evennestdep}.
  \begin{align*}
    \phi_{\circ}(x,y,X_1) =~ \Bigl[x < y \wedge \Bigl(&\exists
    z_1,z_2.~ (z_1
    \leq x \leq y \leq z_2) \wedge \nu(z_1,z_2) \wedge z_1 \in X_1~\wedge \\*
    &\wedge~ \forall z_1', z_2'.~ (z_1 < z_1' \leq x \leq y \leq z_2'
    < z_2) \to \neg \nu(z'_1,z_2') \Bigr)\Bigr]
  \end{align*}
  and let $$\phi_{\leq_2}(x,y,X_1) = ~x=y~ \lor~ (y<x \land
  \phi_{\circ}(y,x,X_1)) ~\lor~ (x<y \land \neg
  \phi_{\circ}(x,y,X_1)).$$ This completes the definition scheme for
  $\Phi_\circ$ which is unambiguous.\eop
\end{proof}

Let $\tau=([n],\lambda,\leq_1,\leq_2)$ be a text. An interval
$[i,j]=\{k \in [n] \mid i \leq_1 k \leq_1j\}$ of the first order is a
\emph{clan} if it is an interval also of the second order.  A
\emph{prime clan} is a clan that does not overlap with any other,
i.e. there is no clan $[k,\ell]$ such that $k<_1i<_1\ell<_1j$ or
$i<_1k<_1j<_1\ell$.

\begin{lemma}\label{lem:trans-primcln}
  Let $nw = (a_1 \ldots a_n, \nu)\in \NW(\Delta)$, let $\Phi_\circ(nw)
  = ([n],\lambda, \leq^\circ_1,\leq^\circ_2)$ and let $
  \Phi_\bullet(nw) = ([n],\lambda,
  \leq^{\bullet}_1,\leq^{\bullet}_2)$. Moreover, let $1 \leq i< j\leq
  n$. \\
  Then $(i,j) \in \nu$ iff $[i,j]$ is a prime clan of
  $\Phi_\circ(nw)$ and we have either $i\ne 1$, $j\ne n$ or $1
  \geq^\circ_2 n$\\\phantom{Then $(i,j) \in \nu$ }iff $[i,j]$ is a prime clan of $\Phi_\bullet(nw)$
  and we have either $i\ne 1$, $j\ne n$ or $1 \leq^\bullet_2 n$.
\end{lemma}
\begin{proof}
  The proof is again by induction on $|\nu|$. If $\nu=\emptyset$, then
  $[1,n]$ is the only prime clan of both $\Phi_\circ(nw)$ and
  $\Phi_\bullet(nw)$ (since any other clan can be overlapped) and we
  have $1 \leq_2^{\circ} n$ and $1 \geq_2^{\bullet} n$.  Now let
  $|\nu|\geq 1$ and let $(i_1,j_1), (i_2,j_2), \ldots, (i_t,j_t)$ with
  $i_1 < i_2 < \ldots < i_t$ be the sequence of surface arches (see
  definition after Def.~\ref{def:nw}).  By definition we have
  \begin{align*}
    \Phi_\circ(nw)~=~\Phi_\circ(nw[1,&i_1-1]) ~\circ~
    \Phi_\circ(nw[i_1,j_1])~ \circ~\cdots~\circ
    \\*
    &~ \circ~ \Phi_\circ(nw[j_{t-1}+1,i_t-1]) ~\circ~
    \Phi_\circ(nw[i_t,j_t]) ~\circ~ \Phi_\circ(nw[j_t+1,n]),
  \end{align*}
  where we ignore a factor if the corresponding interval is empty. We
  show that $(i,j) \in \nu$ iff $[i,j]$ is a prime clan of
  $\Phi_\circ(nw)$ and we have either $i\ne 1$, $j\ne n$ or $1
  \geq^\circ_2 n$. That this holds iff $[i,j]$ is a prime clan of
  $\Phi_\bullet(nw)$ and we have either $i\ne 1$, $j\ne n$ or $1
  \leq^\bullet_2 n$ can again be shown analogously.
  
  (\emph{Only if}).~Let $(i,j) \in \nu$.  Then there is some $r$ such
  that $i_r \leq i < j \leq j_r$.

  If $i=i_r$ or $j=j_r$, then $i=i_r$ and $j=j_r$.  Clearly,
  $[i_r,j_r]$ is a clan.  Suppose for contradiction that there is a
  clan $[\ell,k]$ overlapping $[i_r,j_r]$.  Assume $\ell<i_r< k<j_r$ (the
  case $i_r< \ell<j_r<k$ is similar). By definition of $\Phi_\circ$ we
  get $\ell \leq_2^{\circ} j_r \leq_2^\circ i_r$. 
  Contradiction. Thus $[i_r,j_r]$ is a prime clan.  In particular if
  $i_r=1$ and $j_r=n$, we get $1 \geq_2^\circ n$.

  Otherwise, in case of $i_r < i<j<j_r$, the interval $[i-i_r,j-i_r]$ is a
  prime clan of ${\Phi_\bullet(nw[i_r+1,j_r-1])}$ by induction
  hypothesis. Thus, $[i,j]$ must be a clan, since $[i_r,j_r]$ is a
  clan, too.  Suppose for contradiction that there is a clan $[\ell,k]$
  overlapping $[i,j]$. As $[i-i_r,j-i_r]$ is a prime clan of
  $\Phi_\bullet(nw[i_r+1,j_r-1])$ we get either $\ell \leq i_r$ or $k\geq
  j_r$.  Assume $\ell\leq i_r$ (the other case is similar). Now, if
  $\ell<i_r$, we can argue as above and separate $\ell$ and $i_r$.
  Contradiction. If $i_r=\ell$ and $i_r+1<i$, then $[1,k-i_r]$ is a clan
  in $\Phi_\bullet(nw[i_r+1,j_r-1])$ which overlaps $[i-i_r,j-i_r]$.
  Contradiction. And if $\ell=i_r$ and $i_r+1=i$, we get by definition $i
  \leq_2^{\circ} j \leq_2^\circ i_r$.  Again contradiction.  Thus
  $[i,j]$ must be a prime clan.

  (\emph{If}).~Let $[i,j]$ be a prime clan such that not $i=1$, $j=n$
  and $1 \leq_2^{\circ} n$. If $i=1$ and $j =n$, then $1 \geq_2^\circ
  n$ and $(i,j) \in \nu$ by definition of $\Phi_\circ$.  Now suppose
  $1<i$ or $j<n$.  The following intervals (provided they exist) can
  easily seen to be clans: $[1,i_1-1]$, $[i_1,j_1]$, $[j_1+1,n]$,
  $[1,j_1]$ and $[\ell,n]$ for any $\ell\leq i_1$.  From this we conclude
  that either $i_1 \leq i < j\leq j_1$ or $j_1<i$ since otherwise one
  of the clans above would overlap $[i,j]$. If $i=i_1$ or $j=j_1$ then
  $i=i_1$ and $j=j_1$, since $[i_1,j_1-1]$ and $[i_1+1,j_1]$ are
  clans, and hence $(i,j) \in \nu$.  In the case where $i_1 < i < j<
  j_1$, we get that $[i-i_1,j-i_1]$ must be a prime clan of
  $\Phi_\bullet(nw[i_1+1,j_1-1])$ and if $j_1<i$, we get that
  $[i-j_1,j-j_1]$ must be a prime clan of $\Phi_\circ(nw[j_1+1,n])$.
  Hence, in both cases $(i,j) \in \nu$ by induction hypothesis.\eop
\end{proof}

It is not hard to see that the domains of the partial functions
$\Phi^{-1}_\circ$ and $\Phi^{-1}_\bullet$ are $\FO$-definable. Hence,
by the last lemma there is a definition scheme without parameters
consisting of $\FO$-formulae which defines $\Phi^{-1}_\circ$ (or
alternatively $\Phi^{-1}_\bullet$).

\begin{corollary}\label{cor:nwembedd-phi-1def}
  The partial functions $\Phi_\circ^{-1}$ and $\Phi_\bullet^{-1}$ are
  unambiguously $\FO$-definable.
\end{corollary}

So far we have seen that we can translate a formula over nested words
into a formula over texts (and vice versa) such that the formulae
correspond to each other with respect to $\Phi_\circ$
resp. $\Phi_\bullet$.  We will now show that also WPA can simulate
WNWA (and vice versa) with respect to $\Phi_\circ$
resp. $\Phi_\bullet$.

\begin{proposition}\label{prop:reg-spb-nw}
  Let $S:\TXT(\Delta)\to \K$ be regular. Then
  $\Phi_\circ^{-1}(S),\Phi_\bullet^{-1}(S): \NW(\Delta) \to \K$ are
  regular.
\end{proposition}
\begin{proof}
  We show that $\Phi_\circ^{-1}(S)$ is regular. Analogously one can
  show that $\Phi_\bullet^{-1}(S)$ is regular.  Let
  $\PA=(\HS,\VS,\Omega,\mu,\muop,\mucl,\lambda,\gamma)$ be a WPA such
  that $\behave{\PA}=S$.  We construct a WNWA
  $\A=(Q,\iota,\delta,\kappa)$ with state space $Q= (\HS \disjoint \V)
  \times (\Omega \disjoint \{i\})$ such that for all $h_0,h_n \in
  \HS$, $v_0,v_n \in \V$ and $\omega \in \Omega \disjoint \{i\}$ we
  have
  \begin{align}\label{eq:1}
    \sum_{r: (h_0,\omega) \stackrel{nw}{\longrightarrow}
      (h_n,\omega)}\hspace{-2em} \weight_{\A}(r) & =
    \sum_{\substack{r: h_0 \stackrel{\Phi_\circ(nw)}{\longrightarrow}
        h_n}}\hspace{-1em} \weight_{\PA}(r) &\text{ and } \sum_{r:
      (v_0,\omega) \stackrel{nw}{\longrightarrow}
      (v_n,\omega)}\hspace{-2em} \weight_{\A}(r) = \sum_{\substack{r:
        v_0 \stackrel{\Phi_\bullet(nw)}{\longrightarrow}
        v_n}}\hspace{-1em} \weight_{\PA}(r).
  \end{align}
  
  Intuitively, in the first component one simulates the states of the
  WPA and in the second component one stores the most recent open
  bracket. This has to be updated when reading a return position using
  the look-back ability of the WNWA.  We give now the formal
  definition of the transition functions. We give it only on certain
  subsets of their domains. In all other cases we set the values to
  $0$.  Let $a \in \Delta$, $h_1,h_2 \in \HS$, $v_1,v_2 \in \V$,
  $\omega_1 \in \Omega \disjoint \{i\}$ and $\omega_2 \in
  \Omega$. Define
  \begin{align*}
    \deltaint((h_1,\omega_1),a,(h_2,\omega_1)) & = \mu(h_1,a,h_2) \\
    \deltaint((v_1,\omega_1),a,(v_2,\omega_1)) & = \mu(v_1,a,v_2) \\
    \deltacall((h_1,\omega_1),a,(v_1,\omega_2)) & = \sum_{v \in \V}
    \muop(h_1,(_{\omega_2},v) \cdot\mu(v,a,v_1) \\
    \deltacall((v_1,\omega_1),a,(h_1,\omega_2)) & = \sum_{h \in \HS}
    \muop(v_1,(_{\omega_2},h)\cdot \mu(h,a,h_1) \\
    \deltaret((h_1,\omega_2),(v_1,\omega_1),a,(v_2,\omega_1)) & =
    \sum_{h \in \HS}
    \mu(h_1,a,h)\cdot \mucl(h,)_{\omega_2},v_2)  \\
    \deltaret((v_1,\omega_2),(h_1,\omega_1),a,(h_2,\omega_1)) & =
    \sum_{v \in \V} \mu(v_1,a,v)\cdot \mucl(v,)_{\omega_2},h_2).
  \end{align*}
  Observe that for any $nw \in \NW(\Delta)$ and any run $r:q_0
  \stackrel{nw}{\longrightarrow} q_n$ of $\A$ such that
  $\weight_{\A}(r) \ne 0$ the second components of $q_0$ and $q_n$
  coincide and the first components are either both in $\HS$ or both
  in $\VS$.

  Let $nw=(a_1\ldots a_n,\nu)$.  We show Equation~\ref{eq:1} by
  induction on $|\nu|$.  First let $\nu = \emptyset$. Then for all
  $h_0, h_n \in \HS$ and $\omega \in \Omega \disjoint \{i\}$ we have
  \begin{align*}
    \sum_{r: (h_0,\omega) \stackrel{nw}{\longrightarrow} (h_n,\omega)}
    \weight_{\A}(r) & = \sum_{h_1, \ldots, h_{n-1} \in
      \HS}\prod_{j=1}^n \deltaint
    ((h_{j-1},\omega),a_j,(h_j,\omega))=\\& = \sum_{h_1, \ldots
      ,h_{n-1} \in \HS} \prod_{j=1}^n \mu(h_{j-1},a_j,h_j) =
    \sum_{\substack{r: h_0 \stackrel{\Phi_\circ(nw)}{\longrightarrow}
        h_n}} \weight_{\PA}(r).
  \end{align*}
  Similarly we get the claim for $\Phi_\bullet$. Now, let $\nu \ne
  \emptyset$, let $k$ be the minimal call position and let $\ell$ be the
  corresponding return position.  Let $nw_1=nw[1,k-1]$,
  $nw_2=nw[k+1,\ell-1]$ and $nw_3=nw[\ell+1,n]$ (we assume that all nested
  words exist, the cases where they do not exist are similar). Then
  for all $h_0, h_n \in \HS$ and $\omega \in \Omega \disjoint \{i\}$
  we have
  \begin{align*}
    &\sum_{r: (h_0,\omega) \stackrel{nw}{\longrightarrow}
      (h_n,\omega)}
    \weight_{\A}(r) = \\ &\hspace{2em}=  \sum_{\substack{h_{k-1},h_\ell \in \HS \\
        v_k,v_{\ell-1} \in \V \\ \omega_1 \in \Omega} }\sum_{r_1:
      (h_0,\omega) \stackrel{nw_1}{\longrightarrow} (h_{k-1},\omega)}
    \weight_{\A}(r_1)\cdot
    \deltacall((h_{k-1},\omega),a_k,(v_k,\omega_1)) ~\cdot\\*
    &\mbox{}\hspace{5em} \cdot \hspace{-2em}\sum_{r_2:
      (v_{k},\omega_1) \stackrel{nw_2}{\longrightarrow}
      (v_{\ell-1},\omega_1)}\hspace{-2em} \weight_{\A}(r_2)\cdot
    \deltaret((v_{\ell-1},\omega_1),(h_{k-1},\omega),a_{\ell},(h_\ell,\omega))
    \cdot\hspace{-2em} \sum_{r_3: (h_{\ell},\omega)
      \stackrel{nw_3}{\longrightarrow} (h_{n},\omega)}\hspace{-2em}
    \weight_{\A}(r_3)\\
    &\hspace{2em}= \sum_{\substack{h_{k-1},h_\ell \in \HS \\
        v_k,v_{\ell-1} \in \V \\ \omega_1 \in \Omega} }\sum_{r_1: h_0
      \stackrel{a_1 \circ \ldots \circ a_{k-1}}{\longrightarrow}
      h_{k-1}} \weight_{\PA}(r_1) \cdot \sum_{v \in \V}
    \muop(h_{k-1},(_{\omega_1},v)\cdot \mu(v,a_k,v_k) ~\cdot \\*
    &\mbox{}\hspace{5em} \cdot\hspace{-1em} \sum_{r_2: v_{k}
      \stackrel{\Phi_\bullet(nw_2)}{\longrightarrow}
      v_{\ell-1}}\hspace{-1em} \weight_{\PA}(r_2)\cdot \sum_{v' \in \V}
    \mu(v_{\ell-1},a_l,v') \cdot \mucl(v',)_{\omega_1},h_\ell) \cdot
    \hspace{-1em}\sum_{r_3: h_{\ell}
      \stackrel{\Phi_\circ(nw_3)}{\longrightarrow} h_{n}}\hspace{-1em}
    \weight_{\PA}(r_3) \\
    &\hspace{2em}= \sum_{\substack{r: h_0
        \stackrel{\Phi_\circ(nw)}{\longrightarrow} h_n}}
    \weight_{\PA}(r).
  \end{align*}
  Again, the claim is shown similarly for $\Phi_\bullet$.  This
  concludes the proof of Equation~\eqref{eq:1}.  

  Now consider the WNWA with states $Q'=\{\bottom,?,s,\circ,\bullet\}$
  and transition functions $\deltacall',\deltaint',\deltaret'$ given
  for all $a \in \Delta$ and $p\in Q'\setminus\{\bottom\}$ by
  \begin{align*}
    &\deltacall'(\bottom,a,?)=\deltaint'(\bottom,a,s)=\deltacall'(s,a,\circ)=\deltaint'(s,a,\circ)=
    \deltacall'(?,a,?)=\deltaint'(?,a,?)=\\&
    =\deltaret'(?,p,a,?)=
    \deltaret'(?,\bottom,a,\bullet)=\deltacall'(\bullet,a,\circ)=\deltaint'(\bullet,a,\circ)=
    \deltacall'(\circ,a,\circ)=\\&=\deltaint'(\circ,a,\circ)=\deltaret'(\circ,p,a,\circ)=1.
  \end{align*}
  Set any other values of $\deltacall',\deltaint',\deltaret'$ to $0$
  and let the initial distribution $\iota'$ be given by $\iota'(q')=1$
  if $q'=\bottom$ and $0$ otherwise. Observe that in the case where
  the final distribution $\kappa'$ is given by $\kappa'(q')=1$ if
  $q'=\circ$ and $0$ otherwise, the behavior of the automaton is the
  characteristic series of the set of nested words $nw$ such that
  $\Phi_\circ(nw)$ is a $\circ$-product. We collect such nested words
  in $\NW^\circ$. In the case where the final distribution $\kappa'$
  is given by $\kappa'(q')=1$ if $q'=\bullet$ and $0$ otherwise, the
  behavior of the automaton is the characteristic series of the set of
  nested words $nw$ such that $\Phi_\circ(nw)$ is a
  $\bullet$-product. We collect such nested words in
  $\NW^\bullet$. Finally, in the case where the final distribution
  $\kappa'$ is given by $\kappa'(q')=1$ if $q'=s$ and $0$ otherwise,
  the behavior of the automaton is the characteristic series of the
  set of all singleton nested words, i.e. $\Delta$.

  Now consider the product of this automaton with $\A$ which has
  states $Q \times Q'$ and whose transition functions
  $\deltacall^{\times},\deltaint^{\times},\deltaret^{\times}$ is given
  by letting
  $\deltacall^\times((q,q'),a,(p,p')=\deltacall(q,a,p)\cdot\delta(q',a,p')$
  for all $q,p\in Q$ and $q',p'\in Q'$. If we define the initial and
  final distribution $\iota^\times$ and $\kappa^\times$ by letting for
  all $h \in \HS$ and $\omega \in \Omega$
  \begin{align*}
    \iota^\times((h,i),\bottom) & = \lambda(h) &
    \iota((h,\omega),\bottom) & = \sum_{v \in \V}\lambda(v)\cdot
    \muop(v,(_{\omega},h) \\
    \kappa((h,i),\circ) & = \gamma(h)& \kappa((h,i),\circ) & = \sum_{v
      \in \V} \mucl(h,)_{\omega},v) \cdot\gamma(v),
  \end{align*}
  and in any other case by setting the value to $0$, then the behavior
  of the resulting automaton is $\cha{\NW^\circ}\odot
  \Phi_\circ^{-1}(S)$.  Changing the definitions of
  $\iota^\times,\kappa^\times$ appropriately gives automata with
  behavior $\cha{\NW^\bullet}\odot \Phi_\circ^{-1}(S)$ and
  $\cha{\Delta}\odot \Phi_\circ^{-1}(S)$. The automaton obtained from
  disjoint copies of these three automata has hence the behavior
  $\Phi_\circ^{-1}(S)$.\eop
\end{proof}

\begin{proposition}\label{prop:reg-nw-spb}
  Let $S: \NW(\Delta) \to \K$ be a regular series. Then
  $\Phi_\circ(S), \Phi_\bullet(S): \TXT(\Delta) \to \K$ are regular.
\end{proposition}
\begin{proof}
  Let $\A= (Q, \iota, \delta, \kappa)$ be a WNWA.  We define a WPA
  $\PA=(\HS,\VS,\Omega,\mu,\muop,\mucl,\lambda,\gamma)$ with
  \begin{align*}
    \HS =&
    \{q^{\HS} ~|~q \in Q\} \times (\{c,i\} \disjoint \Delta) &\text{
      and }&&
    \VS = \{q^{\V} ~|~q \in Q\} \times (\{c,i\} \disjoint \Delta)
  \end{align*}
  as well as $\Omega = Q$ such that $(\behave{\PA},\Phi_\circ(nw)) =
  (\behave{\A},nw)$ for all $nw \in \NW(\Delta)$. To prove the result
  for $\Phi_\bullet$ only $\lambda$ and $\gamma$ have to be changed.
 
  Intuitively, in the first component one simulates the states of the
  WNWA, in the second component one either selects whether the next
  transition is a call or an internal transition, or one stores the
  letter to simulate a return position with the next bracket.
  Look-back behavior is simulated by storing a state in the opening
  bracket and closing it at the appropriate return position.
 
  We formally define $\mu, \muop, \mucl$ as follows. We give the
  definition only on certain subsets of their domains. In all other
  cases we set their values to $0$.
  \begin{align*}
    \mu((q_1^{\HS},i),a,(q_2^{\HS},i)) &= \deltaint(q_1,a,q_2) &
    \mu((q_1^{\V},i),a,(q_2^{\V},i)) &= \deltaint(q_1,a,q_2) \\
    \mu((q_1^{\HS},c),a,(q_2^{\HS},i)) &= \deltacall(q_1,a,q_2) &
    \mu((q_1^{\V},c),a,(q_2^{\V},i)) &= \deltacall(q_1,a,q_2) \\
    \mu((q_1^{\HS},i),a,(q_1^{\HS},a)) &= 1   &  \mu((q_1^{\V},i),a,(q_1^{\V},a)) &= 1\\
    \muop((q_1^{\HS},i),(_{q_1},(q_1^{\V},c)) &= 1&
    \muop((q_1^{\V},i),(_{q_1},(q_1^{\HS},c)) &= 1\\
    \mucl((q_1^{\HS},a),)_{q_2},(q_3^{\V},i)) &=
    \deltaret((q_1,q_2,a,q_3)&
    \mucl((q_1^{\V},a),)_{q_2},(q_3^{\HS},i)) &= \deltaret((q_1,q_2,a,q_3)\\
    \lambda(q_1^{\HS},i) &= \iota(q_1) & \gamma(q_1^{\HS},i) &=
    \gamma(q_1)
  \end{align*}

  We use induction on $nw=(a_1\ldots a_n,\nu) \in \NW(\Delta)$ to show
  that the defined WPA behaves as required.  More precisely we show
  that for all $q_1,q_2 \in Q$
  \begin{align*}
    \sum_{r:(q_1^{\HS},i) \stackrel{\Phi_\circ(nw)}{\longrightarrow}
      (q_2^{\HS},i)} \weight_{\PA}(r) = \sum_{r:q_1
      \stackrel{nw}{\longrightarrow} q_2} \weight_{\A}(r) =
    \sum_{r:(q_1^{\V},i) \stackrel{\Phi_\bullet(nw)}{\longrightarrow}
      (q_2^{\V},i)} \weight_{\PA}(r).
  \end{align*}
  This is easy to see if $\nu = \emptyset$. Let $\nu \ne \emptyset$
  and let $k$ be the minimal call position and let $\ell$ be the
  corresponding return position.  Let $nw_1=nw[1,k-1]$,
  $nw_2=nw[k+1,\ell-1]$ and $nw_3=nw[\ell+1,n]$ (we assume that all nested
  words exist, the cases where they do not exist are similar). Then
  \begin{align*}
    & \sum_{r:(q_1^{\HS},i) \stackrel{\Phi_\circ(nw)}{\longrightarrow}
      (q_2^{\HS},i)} \weight_{\PA}(r) ~= \\ & = \sum_{q_3,q_4,q_5,q_6
      \in Q} \sum_{r_1: (q_1^{\HS},i) \stackrel{a_1\circ \ldots
        \circ a_{k-1}}{\longrightarrow}(q_3^{\HS},i)}\hspace{-2em}
    \weight_{\PA}(r_1) \cdot \muop((q_3^{\HS},i),
    (_{q_3},(q_3^{\V},c)) \cdot
    \mu((q_3^{\V},c),a_k,(q_4^{\V},i)) ~\cdot\\*
    & \hspace{5em} \cdot \sum_{r_2:(q_4^{\V},i)
      \stackrel{\Phi_\bullet(nw_2)}{\longrightarrow}(q_5^{\V},i)}\hspace{-2em}
    \weight_{\PA}(r_2) \cdot \mu((q_5^{\V},i),a_\ell,(q_5^{\V},a_\ell)) \cdot
    \mucl((q_5^{\V},a_\ell),)_{q_3},(q_6^{\HS},i)) ~ \cdot \\*
    & \hspace{5em}\cdot \sum_{r_3:(q_6^{\HS},i)
      \stackrel{\Phi_\circ(nw_3)}{\longrightarrow}(q_2^{\HS},i)}\hspace{-2em} \weight_{\PA}(r_3)\\
    & = \sum_{q_3,q_4,q_5,q_6 \in Q} \sum_{r_1: q_1
      \stackrel{nw_1}{\longrightarrow}q_3} \weight_{\A}(r_1) \cdot
    \deltacall(q_3,a_k,q_4) \cdot \sum_{r_2:q_4
      \stackrel{nw_2}{\longrightarrow}q_5} \weight_{\A}(r_2) \cdot
    \deltaret(q_5,q_3,a_\ell,q_6) ~\cdot \\* &\hspace{25em} \cdot
    \sum_{r_3:q_6 \stackrel{nw_3}{\longrightarrow} q_2}
    \weight_{\A}(r_3) \\
    & = \sum_{r:q_1 \stackrel{nw}{\longrightarrow} q_2}
    \weight_{\A}(r). 
  \end{align*}
  We can proceed analogously for $\Phi_\bullet$. Now the result
  follows from the definition of $\lambda$ and $\gamma$.  \eop
\end{proof}

We can now prove Theorem~\ref{thm:nw-main}

\begin{proof}[Proof of Theorem~\ref{thm:nw-main}]
  We prove Theorem~\ref{thm:nw-main}(a).  Let $S: \NW(\Delta) \to \K$
  be regular.  By Proposition~\ref{prop:reg-nw-spb}, $\Phi_\circ(S):
  \TXT(\Delta) \to \K$ is regular and hence $\sREMSO(\K)$-definable by
  Theorem~\ref{thm:txt-srmso}. Now we get that
  $\Phi_\circ^{-1}(\Phi_\circ(S))=S$ is $\sREMSO(\K)$-definable by
  Proposition~\ref{prop:deftrans-weightdef} and
  Corollary~\ref{cor:nwspb-phi-def}.
  
  Conversely, let $S: \NW(\Delta) \to \K$ be $\sRMSO(\K)$-definable.
  By Corollary~\ref{cor:nwembedd-phi-1def} and
  Proposition~\ref{prop:deftrans-weightdef}, $\Phi_\circ(S):
  \TXT(\Delta) \to \K$ is $\sRMSO(\K)$-definable and thus by
  Theorem~\ref{thm:txt-srmso} regular. From
  Proposition~\ref{prop:reg-spb-nw} we conclude that
  $\Phi_\circ^{-1}(\Phi_\circ(S))=S$ is regular, too.
  
  Similarly we get Theorem~\ref{thm:nw-main}(b) from
  Theorem~\ref{thm:txt-main}(b). Theorem~\ref{thm:nw-main}(c) follows
  from Theorem~\ref{thm:txt-main}(c).\eop
\end{proof}

Again note that all proofs are constructive.  Hence, given a sentence
$\phi$ in $\sRMSO(\K)$ (resp. $\swRMSO(\K)$, $\MSO(\K)$) we can
effectively construct a WNWA $\A$ such that
$\behave{\A}=\seman{\phi}$. Conversely, given a WNWA $\A$ we can
construct an $\sREMSO(\K)$ sentence $\phi$ such that
$\behave{\A}=\seman{\phi}$. The following results follow now easily
form the corresponding results for series over alternating
texts~\cite{Mat09b}.

\begin{corollary}
  Let $\K$ be a locally finite semiring or let $\K$ be a ring and let
  $S:\NW(\Delta) \to\K$ be regular such that $S(\NW(\Delta))\subseteq
  \K$ is finite. Moreover, let $A \subseteq \K$. Then $S^{-1}(A)$ is
  regular.
\end{corollary}

\begin{corollary}\label{cor:nw-dec}
  Let $\K$ be a computable field or a computable locally finite
  semiring and let $S_1,S_2:\NW(\Delta)\to \K$ be regular.  It is
  decidable whether $S_1=S_2$.
\end{corollary}

\begin{corollary}\label{cor:nw-maletti}
  Let $\K$ be a computable zero-sum free semiring and let
  $S:\NW(\Delta)\to\K$ be regular. It is decidable whether $(S,nw)=0$
  for all $nw\in \NW(\Delta)$.
\end{corollary}

Note that one motivation of transforming formulae in automata is
solving their satisfiability problem. The last two corollaries can be
seen as a extension of this: We have shown that given a formula in
$\phi\in \sRMSO(\K)$ (resp. $\phi\in\swRMSO(\K)$,
resp. $\phi\in\MSO(\K)$) we can effectively translate it into a
weighted nested word automaton $\A$. Now, provided the semiring is
either zero-sum free or locally finite or a field, using the last two
corollaries we can test whether there is a nested word $nw$ which gets
assigned a non-zero value, i.e. $(\behave{\A},nw)=(\seman{\phi},nw)\ne
0$.

\section{An Application to Algebraic Formal Power Series}
\label{cha:algebraic}

In this section we consider algebraic formal power series and show
that they arise as the projections of regular nested word series and
regular alternating text series.  Applying then our logical
characterizations of the latter we obtain characterizations of
algebraic formal power series in terms of weighted logics generalizing
results of Lautemann, Schwentick and Th{\'e}rien \cite{Lauetal94} on
context-free languages.  Algebraic formal power series have been
considered initially already by Chomsky and Sch\"utzenberger
\cite{ChoSch63} and have since been intensively studied by Kuich and
others. Textbooks containing several aspects of algebraic formal power
series are \cite{SalSoi78} and \cite{KuiSal86}. The reader is also
referred to the survey articles \cite{Kui97} and \cite{PetSal08}.

Let $\Delta^*$ be the free monoid over $\Delta$ and let $\epsilon$
denote the empty word.  A formal power series is a function $S:
\Delta^* \to \K$. We denote the empty word by $\epsilon$.  Given two
formal power series $S_1$, $S_2$, their \emph{Cauchy product}, denoted
$S_1 \cdot S_2$ or $S_1S_2$, is given by $(S_1 \cdot S_2,
w)=\sum_{w_1w_2=w} (S_1,w_1)(S_2,w_2)$ for all $w \in \Delta^*$. By
$S_1 \odot S_2$ we denote the pointwise product also called the
\emph{Hadamard product} and by $S_1 +S_2$ their pointwise sum.
Moreover, if $k \in \K$, then the formal power series $k.S$ is given
by $(k.S,w)=k\cdot (S,w)$ for all $w \in \Delta^*$.  Let $\cha{L}$
denote the characteristic series of a language $L \subseteq
\Delta^*$. We identify $w$ and $\cha{\{w\}}$.  Let $\cX$ be an
alphabet of variables such that $\Delta \cap \cX=\emptyset$. A
polynomial $P$ over $(\Delta \cup \cX)$ is a mapping $P: (\Delta \cup
\cX)^* \to \K$ such that its support is finite, i.e. the set $\supp(P)
= \{w \in (\Delta \cup \cX)^* ~|~(P,w) \ne 0 \}$ is finite.

\begin{definition}\label{def:alg-system}
  A collection of polynomials $(P_X)_{X \in \cX}$ over $(\Delta \cup
  \cX)$ is called an \emph{algebraic system} with variables in $\cX$.
\end{definition}

The supports of the polynomials $P_X$ in the last definition are thus
finite sets consisting of words of the form $u_1X_{1} \ldots
u_kX_{k}u_{k+1}$ where $u_j \in \Delta^*$ and $X_{j} \in \cX$.  We say
that a collection $(S_X)_{X \in \cX}$ of formal power series $S_X:
\Delta^* \to \K$ is a \emph{solution} of the algebraic system
$(P_X)_{X \in \cX}$ if for all $X \in \cX$,
$$
S_X = \sum_{u_1X_{1}\ldots u_kX_{k}u_{k+1} \in
  \supp(P_X)}(P_X,u_1X_{1} \ldots u_kX_{k}u_{k+1}). u_1 S_{X_1} \cdots
u_k S_{X_k}u_{k+1}.
$$

An algebraic system $(P_X)_{X\in \cX}$ is \emph{proper} if
$(P_X,Y)=(P_X,\epsilon)=0$ for all $X,Y \in \cX$. A formal power
series $S$ having the property that $(S,\epsilon)=0$ is called
\emph{quasiregular}.  A proper algebraic system has a unique
quasiregular solution~\cite{SalSoi78}, more precisely a proper
algebraic system has exactly one solution $(S_X)_{X\in\cX}$ such that
$(S_X,\epsilon)=0$ for all $X \in \cX$.

\begin{definition}\label{def:algseries}  A formal power series $S:\Delta^* \to \K$ is an
  \emph{algebraic formal power series} if it is a component of the
  quasiregular solution of a proper algebraic system.
  \footnote{
    This definition is given in~\cite{SalSoi78}. In
    \cite{KuiSal86,Kui97} a series $S$ is called algebraic if its
    quasiregular part $ \cha{\Delta^+}\odot S$ is the component of the
    quasiregular solution of a proper algebraic system.
  }
\end{definition}
We note that over the 2-valued Boolean algebra $\Bo$ these series
correspond exactly to the $\epsilon$-free context-free languages. The
bijection is given by $\supp$.

To warm up let us discuss some easy manipulations of algebraic
systems. For this, let us consider some algebraic system $(P_{X})_{X
  \in \cX}$. Let $X,Y\in \cX$.  Clearly, it follows directly from the
definition of a solution that we can substitute an occurrence of $Y$
in some word of the support of $P_X$ by $P_Y$ without altering the
solutions of the system. More formally: Let $u Y v\in \supp(P_X)$.
Let $(P'_X)_{X \in \cX}$ be given from $(P_X)_{X \in \cX}$ by
replacing $P_X$ with the polynomial
$$
\Big(\cha{\supp(P_X)\setminus\{uYv\}} \odot P_X \Big) + (P_X,uYv).uP_{Y}v. 
$$
Then $(P_X)_{X\in \cX}$ and $(P'_X)_{X \in \cX}$ are
\emph{equivalent}, i.e.  any solution of $(P_X)_{X \in \cX}$ is a
solution of $(P'_{X})_{X \in \cX}$ and vice versa.  An algebraic
system $(P_X)_{X \in \cX}$ is called \emph{weakly strict}, if
$\supp(P_X)\subseteq \{\epsilon\} \cup \Delta(\Delta \cup \cX)^*$ for
all $X \in \cX$. Let us now assume that $(P_X)_{X \in \cX}$ is weakly
strict. Then for any fixed $k \in \N$ by repeated substitution we can
obtain an equivalent algebraic system $(P^k_X)_{X \in \cX}$ such that
for all $X \in \cX$ any $w \in \supp(P^k_X)\setminus \Delta^*$
contains at least $k$ letters from $\Delta$. We conclude that any
weakly strict algebraic system $(P_X)_{X \in \cX}$ has a unique
solution $(S_X)_{X \in \cX}$ which is given by $(S_X,w)=(P_X^k,w)$ for
all $w\in \Delta^*$ such that $|w|< k$.

Now, we continue by manipulating $(P^k_X)_{X \in \cX}$.  Let again $X
\in \cX$ and let $w \in \supp(P^k_X)$ with $|w| < k$.  Let $Y\in
\cX\setminus\{X\}$. For any possible choice of occurrences of $X$ in
the support of $P_Y$ we substitute these occurrences by $w$.  More
precisely, for all $Y \in \cX\setminus\{X\}$ replace $P^k_Y$ by the
polynomial
\begin{align*}
  \sum_{\substack{i\in \N \\
      u_1,u_2,\ldots,u_i,u_{i+1}\in(\Delta\cup\cX)^*}}\hspace{-2.5em}
  (P^k_Y,u_1Xu_2\ldots u_iXu_{i+1}). &u_1\cdot(P^k_X,w).w \cdot
  u_2\cdots u_i \cdot (P^k_X,w).w \cdot u_{i+1}.
\end{align*}
Furthermore, replace $P^k_X$ by the polynomial
\begin{align*}
  &\cha{(\Delta\cup\cX)^*\setminus\{w\}}~\odot\\*
  &\sum_{\substack{i\in \N \\
      u_1,u_2,\ldots,u_i,u_{i+1}\in(\Delta\cup\cX)^*}}\hspace{-2.5em}
  (P^k_X,u_1Xu_2\ldots u_iXu_{i+1}).  u_1\cdot (P^k_X,w).w \cdot
  u_2\cdots u_i \cdot (P^k_X,w).w \cdot u_{i+1}.
\end{align*}
Observe that these sums are in fact finite and note that in these
definitions the factors
$u_1,u_2,\ldots,u_i,u_{i+1}\in(\Delta\cup\cX)^*$ may contain
occurrences of $X$. The resulting system is again weakly strict and
has thus a unique solution $(S'_X)_{X\in \cX}$. A straightforward but
cumbersome calculation, which we omit here, shows, using the
distributivity of the semiring of formal power series, that $S'_Y=S_Y$
for all $Y \in \cX\setminus \{X\}$ and
$S'_X=\cha{(\Delta\cup\cX)^*\setminus\{w\}}\odot S_X$.  For fixed
$0\leq k'<k$ by repeated application we can thus obtain a proper and
weakly strict algebraic system $(R_X)_{X \in \cX}$ such that the
quasiregular and unique solution $(T_X)_{X \in \cX}$ is given by
$(\cha{\{w~|~k'< |w|\}}\odot S_X)_{X \in \cX}$.  In particular, it
follows that the quasiregular part $\cha{\Delta^+}\odot S_X$ of $S_X$
is algebraic for any $X \in \cX$.

\subsection{Nested Word Series and Their Projections} 
Next, we consider the projections of regular nested word series and
show that they give rise exactly to the algebraic series. The
projection $\pi(nw)$ of a nested word $nw = (w,\nu) \in \NW(\Delta)$
is simply the word $w$, i.e. we forget the nesting relation. This
projection is canonically generalized to languages $L \subseteq
\NW(\Delta)$ by setting $\pi(L) = \{\pi(nw) ~|~nw \in L\}$ and to
series $S: \NW(\Delta) \to \K$ by letting
\begin{align*}
  \pi(S) : \Delta^* &\to \K \\*
  w &\mapsto \sum\limits_{\substack{nw \in \NW(\Delta)\\ w = \pi(nw)}}
  (S,nw).
\end{align*} 

\begin{proposition}\label{prop:nwproj2algebraic}
  Let $S: \NW(\Delta) \to \K$ be regular. Then $\pi(S):\Delta^* \to
  \K$ is an algebraic formal power series.
\end{proposition}
\begin{proof}
  Let $\A = (Q, \iota,\delta,\kappa)$ be a WNWA such that
  $\behave{\A}=S$.  We define a weakly strict algebraic system
  $(P_{(q_1,q_2)})_{q_1,q_2 \in Q}$ with variables in $Q^2$ such that
  for its solution $(S_{(q_1,q_2)})_{q_1,q_2 \in Q}$ we have for all
  $w \in \Delta^*$ with $|w| \geq 1$:
  \begin{align}\label{eq:ctxfree1}
    (S_{(q_1,q_2)},w) = \sum_{\substack{nw \in \NW(\Delta) \\  \pi(nw)=w}}\sum_{r:q_1
      \stackrel{nw}{\to}q_2} \weight_{\A}(r). 
  \end{align}

  The idea is to simulate the transitions of a weighted nested word
  automaton. For this we will partition the set of nested words of
  length at least two in three different classes. First the class of
  nested words where the first and the last position are either
  corresponding call and return positions or both internal positions.
  The second class consists of nested words where either the first
  position is a call position and the last position is an internal
  position or the last position is a return position and the first
  position is an internal position. And the last class consists of any
  other, i.e. where the first position is a call position and the last
  position is a return position which do not correspond to each other.
  Using this partition we define for all $q_1,q_2 \in Q$ the
  polynomial $P_{(q_1,q_2)}: (\Delta \cup Q^2)^* \to \K$ as follows:
  \begin{align*}
    (P&_{(q_1,q_2)},w) =\\*
    &\begin{cases}
      1 & \text{if } q_1=q_2 \text{ and } w =\epsilon \\
      \deltaint(q_1,a,q_2) & \text{if } w = a \text{ for some }a \in\Delta \\
      \deltaint(q_1,a,q_3) \cdot \deltaint(q_4,b,q_2) ~+ & \text{if }
      w=a(q_3,q_4)b\\
      \hfill \deltacall(q_1,a,q_3) \cdot \deltaret(q_4,q_1,b,q_2) &
      \hspace{2em}\text{for some }a,b \in \Delta, q_3,q_4\in Q
      \\ 
      \deltacall(q_1,a,q_3) \cdot \deltaret(q_4,q_1,b,q_5) \cdot
      \deltaint(q_6,c,q_2) ~+ & \text{if }
      w=a(q_3,q_4)b(q_5,q_6)c \\
      \hfill \deltaint(q_1,a,q_3) \cdot \deltacall(q_4,b,q_5) \cdot
      \deltaret(q_6,q_4,c,q_2) & \hspace{2em}\text{for some }a,b,c \in
      \Delta \\ & \hspace{2em}\text{and } q_3,q_4,q_5,q_6\in Q\\ 
      \deltacall(q_1,a,q_3) \cdot \deltaret(q_4,q_1,b,q_5)~\cdot &
      \text{if } w=a(q_3,q_4)b(q_5,q_6)c(q_7,q_8)d \\
      \hfill \deltacall(q_6,c,q_7)\cdot \deltaret(q_8,q_6,d,q_2) &
      \hspace{2em}\text{for some } a,b,c,d \in \Delta \\ &\hspace{2em}
      \text{and }q_3,q_4,q_5,q_6,q_7,q_8\in Q
      \\
      0 & \text{otherwise.} 
    \end{cases}  
  \end{align*}
  This is a weakly strict algebraic system having a necessarily unique
  solution $(S_{(q_1,q_2)})_{q_1,q_2 \in Q}$. We show by induction on
  the length of $w$ that \eqref{eq:ctxfree1} holds. For $|w|=1$ this
  is easy to see. Now let $|w|>1$. Then\vfill\eject 
  \begin{align*}
    (S&_{(q_1,q_2)},w)~= \\  =&
    \smash{\sum_{q_3,q_4}\in Q}\Big[\deltaint(q_1,a_1,q_3) \cdot
    \deltaint(q_4,a_n,q_2) + \deltacall(q_1,a_1,q_3) \cdot
    \deltaret(q_4,q_1,a_n,q_2) \Big]\\*&\hspace{25em}\vspace{-12 pt}\cdot ~
    (S_{(q_3,q_4)},a_2 \ldots a_{n-1})~+\\
    &+\smash{\sum_{2 \leq i \leq n-1} \sum_{q_3,q_4,q_5,q_6\in Q}}
    \Big[\deltacall(q_1,a_1,q_3) \cdot \deltaret(q_4,q_1,a_i,q_5) \cdot
    \deltaint(q_6,a_n,q_2) ~+ \\  & \hspace{11em}+
    \deltaint(q_1,a_1,q_3) \cdot \deltacall(q_4,a_i,q_5) \cdot
    \deltaret(q_6,q_4,a_n,q_2) \Big] \\* & \hspace{16.5em}\cdot~
    (S_{(q_3,q_4)},a_2\ldots a_{i-1}) \cdot (S_{(q_5,q_6)},a_{i+1}\ldots
    a_{n-1})\\
    &+\hspace{-.5em} \sum_{2 \leq i <j \leq n-1}\hspace{-.5em}
    \sum_{\substack{q_3,q_4,\\ q_5,q_6,q_7,q_8\in Q}}\hspace{-1em}
    \deltacall(q_1,a_1,q_3) \cdot \deltaret(q_4,q_1,a_i,q_5) \cdot
    \deltacall(q_6,a_j,q_7)\cdot \deltaret(q_8,q_6,a_n,q_2) \\*
    & \hspace{7.5em} \cdot(S_{(q_3,q_4)},a_2\ldots a_{i-1}) \cdot
    (S_{(q_5,q_6)},a_{i+1}\ldots
    a_{j-1}) \cdot (S_{(q_7,q_8)},a_{j+1}\ldots a_{n-1}) \\
    =& \sum_{q_3,q_4\in Q}\Big[\deltaint(q_1,a_1,q_3) \cdot
    \sum_{\substack{nw \in \NW(\Delta)\\\pi(nw)=a_2 \ldots a_{n-1}}}\sum_{r:q_3
      \stackrel{nw}{\to}q_4} \weight_{\A}(r) \cdot
    \deltaint(q_4,a_n,q_2) ~+\\* & \hspace{5em}+ \deltacall(q_1,a_1,q_3)
    \cdot \sum_{\substack{nw \in \NW(\Delta)\\\pi(nw)=a_2 \ldots a_{n-1}}}\sum_{r:q_3
      \stackrel{nw}{\to}q_4} \weight_{\A}(r)\cdot
    \deltaret(q_4,q_1,a_n,q_2) \Big]
    ~+ \\
    &+ \sum_{2 \leq i \leq n-1} \sum_{\substack{q_3,q_4,\\q_5,q_6 \in Q}}
    \Big[\deltacall(q_1,a_1,q_3) \cdot\hspace{-1em} \sum_{\substack{nw_1 \in \NW(\Delta)\\\pi(nw_1)=a_2 \ldots
      a_{i-1}}}\sum_{r_1:q_3 \stackrel{nw_1}{\to}q_4}
    \weight_{\A}(r_1)\cdot \deltaret(q_4,q_1,a_i,q_5) ~\cdot \\*
    &\hspace{14.5em} \cdot \hspace{-1em} \sum_{\substack{nw_2 \in \NW(\Delta)\\\pi(nw_2)=a_{i+1} \ldots
      a_{n-1}}}\sum_{r_2:q_5 \stackrel{nw_2}{\to}q_6} \weight_{\A}(r_2) \cdot
    \deltaint(q_6,a_n,q_2) ~+ \\* & \hspace{8em}+ \deltaint(q_1,a_1,q_3)
    \cdot\hspace{-1em} \sum_{\substack{nw_1 \in \NW(\Delta)\\\pi(nw_1)=a_2 \ldots a_{i-1}}}\sum_{r_1:q_3
      \stackrel{nw_1}{\to}q_4} \weight_{\A}(r_1)~\cdot \\* &\hspace{9.5em}\cdot
    \deltacall(q_4,a_i,q_5)\cdot\hspace{-2em}
    \sum_{\substack{nw_2\in \NW(\Delta)\\\pi(nw_2)=a_{i+1} \ldots a_{n-1}}}\sum_{r_2:q_5
      \stackrel{nw_2}{\to}q_6} \weight_{\A}(r_2) \cdot
    \deltaret(q_6,q_4,a_n,q_2)
    \Big]~+ \\
    &+ \sum_{2 \leq i <j \leq n-1}
    \sum_{\substack{q_3,q_4,q_5,\\q_6,q_7,q_8\in Q}}
    \deltacall(q_1,a_1,q_3) \cdot\hspace{-1em} \sum_{\substack{nw_1 \in \NW(\Delta)\\\pi(nw_1)=a_{2} \ldots
      a_{i-1}}}\sum_{r_1:q_3 \stackrel{nw_1}{\to}q_4} \weight_{\A}(r_1) \cdot
    \deltaret(q_4,q_1,a_i,q_5)~\cdot \\* &\hspace{15em} \cdot
    \sum_{\substack{nw_2\in \NW(\Delta)\\\pi(nw_2)=a_{i+1} \ldots a_{j-1}}}\sum_{r_2:q_5
      \stackrel{nw_2}{\to}q_6} \weight_{\A}(r_2) ~\cdot \\*
    &\hspace{8em}\cdot \deltacall(q_6,a_j,q_7)\cdot
    \sum_{\substack{nw_3\in \NW(\Delta)\\\pi(nw_3)=a_{j+1} \ldots a_{n-1}}}\sum_{r_3:q_7
      \stackrel{nw_3}{\to}q_8} \weight_{\A}(r_3) \cdot \deltaret(q_8,q_6,a_n,q_2) \\
    =&\sum_{\pi(nw)=w}\sum_{r:q_1 \stackrel{nw}{\to}q_2}
    \weight_{\A}(r). 
  \end{align*}

  Now, let $X$ be a fresh variable and extend
  $(P_{(q_1,q_2)})_{q_1,q_2\in Q}$ by adding the new polynomial
  $P_X=\sum_{q_1,q_2\in Q}\iota(q_1)\cdot\kappa(q_2).P_{(q_1,q_2)}$.
  Clearly, the unique solution of this extended system is obtained by
  adding $S_X=\sum_{q_1,q_2\in
    Q}\iota(q_1)\cdot\kappa(q_2).S_{(q_1,q_2)}$ to
  $(S_{(q_1,q_2)})_{q_1,q_2\in Q}$. The quasiregular part of $S_X$
  equals $\pi(\behave{\A})$ which is thus algebraic by our
  considerations after Definition~\ref{def:algseries}.\eop
\end{proof}

Given an algebraic system $(P_X)_{X \in \cX}$ over $(\Delta \cup \cX)$
and some $X \in \cX$, we define the \emph{underlying grammar}
$G_X=(\Delta, \cX, X,F)$ where the set $F \subseteq \cX
\times(\cX\cup\Delta)^*$ of productions is given by letting $(Y,w) \in
F$ iff $(P_Y,w) \ne 0$. Let $u \in \Delta^*$.  A \emph{derivation
  tree} of $u$ under $G_X$ is a finite tree $t$ such that the
following holds:
\begin{enumerate}[(a)]
\item The root is labeled with $(X,w)$ for some $w \in \supp(P_X)$.
\item For each inner node $v$ with label $(Y,w)$ the first component
  of the labels of the children of $v$ from left to right yield $w$.
\item The labels of the leaves  from left to right yield $u$. 
\end{enumerate}

\noindent We collect all derivation trees $t$ of $u$ under $G_X$ in
$\Der(G_X,u)$.  Clearly, if $(P_X)_{X \in \cX}$ is proper, then each
inner node of $t$ either has a single leaf attached or branches at
least binarily. Hence, in this case $\Der(G_X,u)$ is a finite set.
Let $v$ be a node of $t$. If $v$ is an inner node and $(Y,w)$ its
label, then we let $\weight(t,v)=(P_Y,w)$. If $v$ is a leaf, we let
$\weight(t,v)=1$. Now we define the weight $\weight(t)$ of $t$ by
$\weight(t)=\prod_{v \text{ node of } t} \weight(t,v)$. The following
lemma seems to belong to what is sometimes called folklore, it can
easily be shown by induction on the length of $w$. A proof of a
similar but weaker result can be found in
\cite[Theorem~IV.1.5]{SalSoi78}.

\begin{lemma}\label{lem:sumDerivatTree}
  Let $(P_X)_{X \in \cX}$ be a proper algebraic system and let
  $(S_X)_{X \in \cX}$ be its unique quasiregular solution. Then
  $$
  (S_X,w) = \sum_{t \in \Der(G_X,w)} \weight(t)~~~\text{ for all } X
  \in \cX \text{ and } w \in \Delta^*. 
  $$
\end{lemma}

We now show the converse of Proposition~\ref{prop:nwproj2algebraic}.

\begin{proposition}\label{prop:algserAreProj}
  Let $R: \Delta^* \to \K$ be an algebraic formal power series. Then
  there is a regular nested word series $S: \NW(\Delta) \to \K$ such
  that $\pi(S) =R$.
\end{proposition}
\begin{proof}
  Let $(P_X)_{X \in \cX}$ be a proper algebraic system with
  quasiregular solution $(S_X)_{X \in \cX}$ and let $Y \in \cX$ such
  that $R=S_Y$. We construct a WNWA $\A = (Q, \iota,\delta,\kappa)$
  such that ${\pi(\behave{\A})} = S_Y$. Any element in the support of
  some $P_X$ will define a transition in the automaton. In order not
  to produce $\epsilon$-transitions, we require that each word in the
  support of some $P_X$ contains an element of $\Delta$, and in order
  to produce at most one call for each transition, each word in the
  support of some $P_Y$ contains at most two elements of $\cX$.
  Therefore we assume the algebraic system $(P_X)_{X \in \cX}$ to be
  in Greibach normal form \cite{KuiSal86}, i.e. we require that
  $\supp(P_X) \subseteq \Delta \cup \Delta \cX \cup \Delta \cX\cX$ for
  all $X \in \cX$.  Elements of $\Delta \cX\cX$ produce call
  transitions, elements in $\Delta \cX$ produce internal transitions
  and elements in $\Delta$ produce return transitions.  More
  precisely, let $Q =(\cX\cup \{\bottom\}) \times (\cX \cup
  \{\bottom\})$ for some fresh symbol $\bottom$, and for all
  $X_1,X_3,X_4 \in \cX$ and $X_2 \in \cX \cup \{\bottom\}$ let
  \begin{align*}
    \deltacall((X_1,X_4), a, (X_3,X_2)) &= (P_{X_1},aX_3X_4)\\
    \deltaint((X_1,X_2), a ,(X_3,X_2)) &= (P_{X_1},aX_3) \\
    \deltaret((X_1,X_2),(X_3,X_4), a ,(X_4,X_2)) &= (P_{X_1},a).
  \end{align*}
  Moreover, let $\deltaint((X_1,\bottom),a,(\bottom,\bottom)) =
  (P_{X_1},a)$.  Any other transition gets weight $0$. Furthermore,
  for all $X,Z \in \cX\cup\{\bottom\}$ we let
  \begin{align*}
    \iota(X,Z) = \begin{cases}
      1 & \text{if } X=Y\\
      0 & \text{otherwise}
    \end{cases}
    &&
    \kappa(X,Z) = \begin{cases}
      1 & \text{if } X = Z = \bottom \\
      0 & \text{otherwise}. 
    \end{cases}
  \end{align*}

  The idea is to simulate a derivation tree of the underlying grammar
  $G_Y$  traversed from the left to
  the right.  More precisely, when processing a production
  $(X_1,aX_3X_2)$ in a derivation tree, then a call transition is
  executed and we continue in a state with first component $X_3$. At
  the return position the automaton changes to $X_2$. Since the
  automaton looks back to the state in which the automaton was
  \emph{before} the corresponding call position, it has to guess $X_2$
  in advance.  This is stored in the second component which was
  introduced for this reason.  One can show by induction on $|w|$ that
  for all $w \in \Delta^*$, $X \in \cX$ and $Z'\in \cX \cup
  \{\bottom\}$ we have
  \begin{align}\label{eq:afpsAreProj-1}
    (S_X,wa)=\sum_{\nu \in \Nest_{|w|}}\sum_{\substack{X'\in \cX, Z\in
        \cX \cup\{\bottom\} \\r:(X,Z)\stackrel{(w,\nu)}{\to}(X',Z')}}
    \weight_{\A}(r) \cdot (P_{X'},a)
  \end{align}
  where we make the convention that there is a run
  $r:(X,Z)\stackrel{(\epsilon,\emptyset)}{\to}(X',Z')$ iff $X=X'$ and
  $Z=Z'$. Moreover, for this run we let $\weight_\A(r)=1$.
  Now the result follows easily from the observation that by the
  definition of $\delta$ the last transition of a run
  $r:(Y,Z)\stackrel{(w,\nu)}{\to}(\bottom,\bottom)$ with
  $\weight(r)\ne 0$ must be an internal transition.\eop
\end{proof}

Subsequently we make use of the following well known
result~\cite{KuiSal86}.  We just indicate how it can be obtained in
this context using Propositions~\ref{prop:nwproj2algebraic}
and~\ref{prop:algserAreProj}, but note that a more elementary proof
and more general results can be found in \cite[Chapter 15]{KuiSal86}.

\begin{corollary}[Kuich \& Salomaa\mbox{\cite[Lemma~15.2]{KuiSal86}}]\label{cor:normalform}
  Let $S:\Delta^* \to \K$ be an algebraic formal power series. Then
  there is an algebraic system $(P_X)_{X \in \cX}$ such that
  $\supp(P_X)\subseteq\Delta \cup \Delta(\Delta \cup \cX)^*\Delta$ for
  all $X \in \cX$ and $S=S_X$ for some $X \in \cX$.
\end{corollary}
\begin{proof}
  By Proposition~\ref{prop:algserAreProj}, $S$ is the projection of
  some regular nested word series $R:\NW(\Delta) \to \K$. Now let $\A$
  be a WNWA and $Q$ its set of states such that
  $\behave{\A}=R$. Consider the weakly strict algebraic system
  $(P_X,(P_{(q_1,q_2)})_{q_1,q_2 \in Q})$ of the proof of
  Proposition~\ref{prop:nwproj2algebraic} and its unique solution
  $(S_X,(S_{(q_1,q_2)})_{q_1,q_2 \in Q})$. Using the manipulations
  given after Definition~\ref{def:algseries} we can transform this
  system into a system of the required form having as a solution the
  quasiregular part of $S_X$ which equals $S$.\eop
\end{proof}

\subsection{A Logical Characterization of Algebraic Formal Power
  Series}
Our aim is to give a logical characterization of algebraic formal
power series in the spirit of Lautemann, Schwentick and
Th{\'e}rien~\cite{Lauetal94}. They showed that the context-free
languages are precisely the languages which can be defined by
second-order sentences over words of the form $\exists \nu.  \phi$
where $\phi$ is a first-order formula and $\nu$ a binary predicate
ranging over nesting relations\footnote{In~\cite{Lauetal94} nesting
  relations were named \emph{matchings}.}.  We identify a word
$a_1\ldots a_n \in\Delta^*$ with the structure $([n],\leq,\lambda)$,
where $\leq$ is the canonical order of $[n]$ and $\lambda:[n]\to
\Delta$ is given by $\lambda(i)=a_i$ for all $i\in [n]$.  Let $\phi$
be a weighted second-order formula over words containing, apart from a
single $2$-ary relation variable $\nu$, only $1$-ary relation
variables. In other words let $\phi\in \MSO(\K,\signw{\Delta})$. Let
$\Free(\phi) \subseteq \V$, $w \in \Delta^*$ and $\gamma$ a
$(\V,w)$-assignment.  We define the semantics $\seman{\exists \nu.
  \phi}^{\text{nest}}: \Delta^* \to \K$ by letting
$$(\seman{\exists \nu.  \phi}^{\text{nest}},(w,\gamma)) =
\sum\limits_{\nu \in \Nest_{|w|}} (\seman{\phi},((w,\nu),\gamma)).$$

Using our characterization of nested word automata by means of
weighted logics (Theorem~\ref{thm:nw-main}), we may reformulate
Proposition~\ref{prop:nwproj2algebraic} as follows:

\begin{corollary}\label{cor:matchdef2algebraic}
  Let $\phi \in \sRMSO(\K,\signw{\Delta})$ be a sentence. Then
  $\seman{\exists \nu. \phi}^{\text{nest}}:\Delta^* \to \K$ is an
  algebraic formal power series.
\end{corollary}

Next we show a result which sharpens
Proposition~\ref{prop:algserAreProj}. For this we follow the proof of
Lautemann, Schwentick and Th{\'e}rien \cite[Theorem~2.1]{Lauetal94}
with small changes in the details.

\begin{proposition}\label{prop:algseraredef}
  Let $S$ be an algebraic formal power series. Then there is a
  sentence $\phi \in \sRFO(\K,\signw{\Delta})$ such that $S =
  \seman{\exists \nu.  \phi}^{\text{nest}}$.
\end{proposition}
\begin{proof}
  We use an idea of Lautemann, Schwentick and Th{\'e}rien
  \cite{Lauetal94} and adapt it to the weighted setting.  This
  requires that we have to be more careful in order not to count
  weights twice.

  \emph{A normal form.}  By Corollary~\ref{cor:normalform} we may
  assume that $S$ is the component of the solution of an algebraic
  system with variables in $\cX$ having all supports in $\Delta \cup
  \Delta(\Delta \cup \cX)^*\Delta$.  By the transformations discussed
  after Definition~\ref{def:algseries} we obtain from this a proper
  algebraic system $(P'_{X})_{X \in \cX}$ with solution $(S_{X})_{X
    \in \cX}$ such that for all $X \in\cX$, $\supp(P'_{X})$ does not
  contain elements of $\Delta \cup \{\epsilon\}$ and $\cha{\{w \in
    \Delta^* ~|~|w| >1\}}\odot S$ is a component of the solution.
  Clearly, it suffices to show the proposition for the latter series
  instead of $S$.

  Now we proceed as in \cite{Lauetal94} and transform the system
  $(P'_X)_{X\in \cX}$ into an equivalent system $(P_{X})_{X \in \cX}$.
  Let $w \in \supp(P'_{X})$ for some $X \in \cX$.  The image of $w$
  under the homomorphism which is the identity on $\Delta$ and maps
  any $Y \in \cX$ to the fresh symbol $|$ is called the \emph{pattern}
  $\text{patt}(w)$ of $w$. Let us now fix a strict linear order $<$ on
  $\cX$.  Similarly to \cite{Lauetal94}, we proceed along this linear
  order. Let $X$ be the current symbol. In order to obtain $P_X$ we
  substitute iteratively some $Z \in \cX$ in some $w \in \supp(P'_X)$
  by $P'_{Z}$ (cf.  considerations after
  Definition~\ref{def:algseries}) until for all $Y \in \cX$, with
  $Y<X$, $\text{patt}(w)\ne\text{patt}(w')$ for all $w'\in
  \supp(P_Y)\setminus \Delta^*$ and $w \in \supp(P_X)\setminus
  \Delta^*$.  This is possible since by our considerations after
  Definition~\ref{def:algseries} we can ensure that all elements in
  $\supp(P_X)\setminus \Delta^*$ are longer than all elements in
  $\supp(P_Y)\setminus \Delta^*$ for all $Y < X$.  We finally obtain a
  proper algebraic system $(P_{X})_{X \in \cX}$ equivalent to
  $(P'_X)_{X \in \cX}$ having the following properties:
  \begin{enumerate}[(1)]
  \item $\supp(P_{X}) \subseteq \Delta (\Delta \cup \cX)^+ \Delta$ for
    all $X \in \cX$.
  \item For all $X,Y \in \cX$, if $\text{patt}(w) =\text{patt}(w')$
    for some $w \in \supp(P_{X})\setminus \Delta^*$ and $w' \in
    \supp(P_{Y})\setminus\Delta^*$, then $X=Y$.
  \end{enumerate}
  Let us fix $Y \in \cX$. We now proceed by giving a sentence $\phi_Y
  \in \sRFO(\K)$ such that $\pi(\seman{\phi_Y}) = S_{Y}$. This will
  conclude the proof.

  \emph{Some macros.}  Let $G_Y$ be the underlying grammar (see the
  definition after the proof of
  Proposition~\ref{prop:nwproj2algebraic}) and let $u \in \Delta^*$.
  The basic idea now is to assign to each derivation tree $t \in
  \Der(G_Y,u)$ a nesting relation $\nu_t$ of width $|u|$. This is done
  by letting $(i,j) \in \nu_t$ if there is an inner node of $t$ such
  that the leaves of the subtree rooted at this node are exactly the
  leaves between the $i$th and the $j$th leaf of $t$ (in lexicographic
  order including the $i$th and the $j$th leaf).  Clearly, due to the
  special form of $(P_X)_{X \in \cX}$ this binary relation is indeed a
  nesting relation.  Let us now define some macros for nested words.
  Let $nw=(u,\nu)= (a_1\ldots a_k,\nu)\in \NW(\Delta)$. Then let
  $\min(x)$ and $\max(y)$ express that $x$ is assigned the first
  position and $y$ the last position.  Furthermore, the formula
  $\text{inchild}(x,y)$ express that $(x,y) \in\nu$ corresponds to an
  inner node of $t$ which has an inner node as a child.
  $$
  \text{inchild}(x,y) = \nu(x,y) \wedge \exists z,z'.~ (x<z<y) \wedge
  \nu(z,z')
  $$

  The macro $\text{surf}(x,y,x_1,y_1)$ says that $(x_1,y_1)$ is a
  surface arch of $nw[x,y]$:
  \begin{align*}
    \text{surf}(x,y,x_1,y_1) = (x < x_1< y_1 < y) &\wedge
    \nu(x_1,y_1)~\wedge\\* &\wedge \forall z,z'.~(x < z < x_1 <y_1 < z'
    < y) \to \neg \nu(z,z').
  \end{align*}
  As in \cite{Lauetal94}, for $v \in\Delta^*$ let $\psi_{v}(x,y)$ be a
  first-order formula that expresses there is no call strictly between
  positions $x$ and $y$ and that the substring given by the positions
  strictly between position $x$ and $y$ equals $v$.  For a word $w=avb
  \in \Delta^+$ define $\theta_{w}(x,y)$ as follows.
  \begin{align*}
    \theta_{w}(x,y) = \Lab_a(x) \wedge \Lab_b(y) \wedge \psi_{v}(x,y).
  \end{align*}
  Now we will need the notion of a pattern also for nested words
  \cite{Lauetal94}.  Let $(i_1,j_1), \ldots, (i_s,j_s)$ be the
  sequence of all \emph{surface arches} of $nw$.  The \emph{pattern}
  $\text{patt}(nw)$ of $nw$ is the string $a_1 \ldots a_{i_1-1} ~|~
  a_{j_1+1} \ldots a_{i_s-1}~|~a_{j_s+1} \ldots a_k$.  Now, let $X \in
  \cX$, let $w =av_0X_1v_1 \ldots v_{s-1}X_sv_s b \in \supp(P_{X})
  \setminus \Delta^+$ and let $p=\text{patt}(w)= av_0|v_1 \ldots
  v_{s-1}|v_s b$.  We define the formula $\chi_p(x)$ (cf.
  \cite{Lauetal94}) which states that $x$ is a call position with
  return position $y$ and $\text{patt}(nw[x,y]) =p$.
  \begin{align*}
    \chi_p(x) = &\exists y.~\nu(x,y) \wedge \Lab_{a}(x) \wedge
    \Lab_b(y) ~\wedge \\* & \wedge \exists x_1, y_1, \ldots, x_s,y_s.
    \Big[ (x < x_1< y_1 \ldots <y_s<y) \wedge \psi_{v_0}(x,x_1) \wedge
    \ldots \wedge \psi_{v_s}(y_s,y) ~\wedge \\*
    &\hspace{15.9em} \wedge ( \text{surf}(x,y,x_1,y_1) \wedge \ldots
    \wedge \text{surf}(x,y,x_s,y_s)\Big]
  \end{align*}
  Now let $\widetilde{\chi}_{X}(x)$ be the disjunction of all
  $\chi_p(x)$ over all patterns $p$ of words $w \in \supp(P_{X})
  \setminus \Delta^+$ and let $\widetilde{\theta}_{X}(x,y)$ be the
  disjunction of all $\theta_w(x,y)$ over $w \in \supp(P_{X}) \cap
  \Delta^+$.  Let again $w =av_0X_1v_1 \ldots v_{s-1}X_sv_s b \in
  \supp(P_{X}) \setminus \Delta^+$.  Similarly to \cite{Lauetal94} we
  define now the formula $\widetilde{\chi}_w(x,y)$:
  \begin{align*}
    \widetilde{\chi}_w&(x,y) = \exists y.~\nu(x,y) \wedge \Lab_{a}(x)
    \wedge \Lab_b(y) ~\wedge \\* & \wedge \exists x_1, y_1, \ldots,
    x_s,y_s.  \Big[ (x < x_1< y_1 \ldots <y_s<y) \wedge
    \psi_{v_0}(x,x_1) \wedge
    \ldots \wedge \psi_{v_s}(y_s,y) ~\wedge \\*
    &\hspace{15.5em} \wedge ( \text{surf}(x,y,x_1,y_1) \wedge \ldots
    \wedge \text{surf}(x,y,x_s,y_s)\wedge \\&\hspace{10em} \wedge
    \Big(\widetilde{\chi}_{X_1}(x_1) \lor \theta_{X_1}(x_1,y_1)\Big)
    \wedge \ldots \wedge \Big(\widetilde{\chi}_{X_s}(x_s) \lor
    \theta_{X_s}(x_s,y_s)\Big)\Big].
  \end{align*}
  We show in the next paragraph that there is a bijective
  correspondence between the set of derivation trees $t \in
  \Der(G_Y,u)$ and the nested words $(u,\nu)$ satisfying the following
  formula
  \begin{align*}
    \psi_Y = \exists x,y.~ \min(x) \wedge \max(y) \wedge \nu(x,y)
    &\wedge \Big( \widetilde{\chi}_{Y}(x,y) \lor  \widetilde{\theta}_{Y}(x,y) \Big)~ \wedge\\
    &\wedge~\forall z,z'.~\text{inchild}(z,z')\to\hspace{-1em}
    \bigvee_{\substack{X \in \cX \\w \in \supp(P_{X}) \setminus
        \Delta^+}}\hspace{-1em} \widetilde{\chi}_{w}(z,z').
  \end{align*}  

  \emph{The formula.}  Given a derivation tree $t\in\Der(G_Y,u)$ we
  assign to it a nesting relation $\nu_t$ as described above.
  Clearly, $(1,n)\in \nu_t$ and either
  $(u,\nu_t)\models\widetilde{\theta}_{Y}[1,n]$ or
  $(w,\nu_t)\models\widetilde{\chi}_{Y}[1,n]$. Furthermore, if $1 \leq
  i<j \leq n$ and $(u,\nu_t)\models \text{inchild}[i,j]$, then there
  is an inner node of $t$ such that the leaves of the subtree rooted
  at this node are exactly the leaves between the $i$th and the $j$th
  leaf of $t$. Let $(X,w)$ be the label of this inner node, then
  $(u,\nu_t)\models \widetilde{\chi}_{w}[i,j]$ by construction and
  hence $(u,\nu_t)\models \psi$.  Conversely, let $\nu$ be a nesting
  relation such that $(u,\nu)\models\psi$. We define a derivation tree
  $t_\nu$ inductively as follows.  If $\{(1,n)\}=\nu$, then $t_\nu$
  consists of a single inner node, the root, labeled by $(Y,u)$. In
  this case we must have $(u,\nu)\models \widetilde{\theta}_{Y}[1,n]$
  and hence $t_{\nu}$ is a derivation tree.  Otherwise, let
  $(i_1,j_1), \ldots, (i_s,j_s)$ be the sequence of surface arches of
  $(u,\nu\setminus\{(1,n)\})$ and let $a^1_1\ldots a^1_{n_1}|\ldots
  |a_1^s \ldots a_{n_s}^s |a_1^{s+1}\ldots a_{n_{s+1}}^{s+1}$ be the
  pattern of $(u,\nu\setminus\{(1,n)\})$. Moreover, for $1 \leq k \leq
  s$ let $u[i_k,j_k]$ be the subword of $u$ from the $i_k$th position
  to the $j_k$th position. Then we must have
  $$(u,\nu)\models \bigvee_{\substack{X \in \cX \\w \in \supp(P_{X})
      \setminus \Delta^+}}\hspace{-1em} \widetilde{\chi}_{w}[1,n]$$
  and hence for all $1 \leq k \leq s$ we have $(u,\nu)[i_k,j_k]
  \models \psi_{X_k}$ for some $X_k\in \cX$. Thus by inductions
  hypothesis there are $t_k \in \Der(G_{X_k},u[i_k,j_k])$. 
  We define $t_{\nu}$ to be the tree whose root is labeled $(Y,a^1_1\ldots a^1_{n_1}X_1\ldots
  X_sa_1^{s+1}\ldots a_{n_{s+1}}^{s+1})$ and where the trees rooted at
  the children of the root are as follows from left to right: $a^1_1,\ldots,
  a^1_{n_1},t_1,\ldots, t_s,a_1^{s+1},\ldots, a_{n_{s+1}}^{s+1}$. 
  We conclude that $t_{\nu}$ is a derivation tree, since
  $(u,\nu)\models\widetilde{\chi}_{Y}[1,n]$. 

  Now we can give the formula $\phi_Y$.
  \begin{align*}
    \phi_Y = \psi_Y^+ ~\wedge~ \forall x,y.~ \nu(x,y) \to
    \bigvee_{\substack{X \in \cX\\w \in \supp(P_X)}} \Big(
    &\text{inchild}(x,y) \impl \big(\widetilde{\chi}_w(x,y)^+ \wedge
    (P_X,w)\big) ~\wedge\\* &\wedge~ \neg \text{inchild}(x,y) \impl
    \big(\theta_w(x,y)^+ \wedge (P_X,w)\big)\Big)
  \end{align*}
  Let $t\in \Der(G_Y,u)$ and let $\nu_t$ be the corresponding nesting
  relation. By construction $(\seman{\phi_Y},(u,\nu_t))=\weight(t)$
  and thus $\seman{\exists \nu.  \phi_Y}^{\text{nest}} = S_{Y}$ by
  Lemma \ref{lem:sumDerivatTree}.  \eop
\end{proof}

Let us summarize our results of this section so far.
\begin{theorem}\label{thm:algebraiccharac}
  Let $\K$ be a commutative semiring and let $S: \Delta^* \to \K$ be a
  formal power series. Then the following are equivalent:
  \begin{enumerate}[\em(1)]
  \item $S$ is an algebraic formal power series.
  \item $S = \pi(R)$ for some regular $R:\NW(\Delta) \to \K$. 
  \item There is a sentence $\phi \in \sRFO(\K,\signw{\Delta})$
    such that $\seman{\exists \nu. \phi}^{\text{nest}} =S$. 
  \end{enumerate}
\end{theorem}
\begin{proof}
\noindent{(1) $\Rightarrow$ (3)}.\enspace This is Proposition~\ref{prop:algseraredef}. 

\noindent{(3) $\Rightarrow$ (2)}.\enspace Follows from
Theorem~\ref{thm:nw-main}(a) and the definition of $\pi$. 

\noindent{(2) $\Rightarrow$ (1)}.\enspace This is
Proposition~\ref{prop:nwproj2algebraic}.\eop
\end{proof}

Let $\K = \N$ and let $S: \{a\}^+ \to \N$ be an algebraic series. As
$S=\pi(R)$ for some regular nested word series $R$, it is not hard to
see that $(S,a^n) \leq 2^{n^2}\cdot c^{n}$ for some constant $c$ and
all $n \in \N$. Using weighted pushdown automata (cf. \cite{KuiSal86})
one can even show that $(S,a^n) \leq c^{n}$ for some constant $c$ and
all $n \in \N$. Thus in item 3 of the last result we may not replace
$\sRFO(\K)$ by $\FO(\K)$ since $(\seman{\forall x. \exists y.
  1},a^n)=n^n$. 

Again we note that all proofs are effective and given a proper
algebraic system $(P_X)_{X \in \cX}$ with solution $(S_X)_{X \in \cX}$
and an effectively given semiring $\K$, we can compute an $\sRFO(\K)$
sentence $\phi_Y$ for all $Y \in \cX$ such that $S_Y = \seman{\exists
  \nu.\phi_Y}^{\text{nest}}$ and vice versa.

\subsection{\!Yet Another Characterization of Algebraic Formal Power
  Series}\!\!
Even though our logical characterization of regular nested word series
(Theorem~\ref{thm:nw-main}) might also be obtained by structural
induction, the connection between alternating texts and nested words
we established enables us now to also obtain a generalization of the
second main result of~\cite{Lauetal94}.  In this paper another logical
characterization of context-free languages was given where
quantification over nesting relations is now replaced by
quantification over tree-definable orders. In \cite{Lauetal94} a
linear order $\leq$ on $[n]$ was called tree-definable if there is a
binary tree $t$ with $n$ leaves which are labeled $1, \ldots,n$ in
lexicographic order and whose internal nodes are labeled with
$\{\swarrow,\searrow\}$ such that $i \leq j$ iff $i$ is visited before
$j$ in the depth-first traversal of $t$ in which, at every node with
label $\swarrow$, first the left, and at every node with label
$\searrow$, first the right child is visited.  We will give a slightly
different definition which is easily seen to be equivalent by simply
replacing $\swarrow$ by $\bullet$ and $\searrow$ by $\circ$.

\begin{definition}\label{def:tree-def}
  Let $n \in \N_+$ and let $\leq_1$ be the canonical order of
  $[n]$. Moreover, let $\lambda:[n] \to \Delta$ be a labeling.  A
  linear order $\leq_2$ of $[n]$ is \emph{tree-definable} iff
  $([n],\lambda,\leq_1,\leq_2)$ is an alternating text.
\end{definition}
We collect all tree-definable orders of $[n]$ in $\TDO_n$.  Our aim is
now to extend the above mentioned result of \cite{Lauetal94} and to
show, using the connection between nested words and alternating texts,
that a formal power series is an algebraic formal power series iff it
can be defined by a second-order sentence over words of the form
$\exists \leq_2.  \phi$ where $\phi$ is a first-order formula and
$\leq_2$ a binary relation symbol ranging over tree-definable orders.
Note that like matchings, tree-definable orders are first-order
definable relations\cite{HooPas97,Mat09b}.  First, we start by
defining the projection $\pi(\tau)$ of an alternating text
$\tau=([n],\leq_1,\leq_2,\lambda) \in \TXT(\Delta)$ to be the word
$([n],\leq_1,\lambda)$, i.e. we forget the second order. As for nested
words, this projection is canonically generalized to languages $L
\subseteq \TXT(\Delta)$ by setting $\pi(L) = \{\pi(\tau) ~|~\tau \in
L\}$ and to series $S: \TXT(\Delta) \to \K$ by letting
\begin{align*}
  \pi(S) : \Delta^* &\to \K \\*
  w &\mapsto \sum_{\substack{\tau \in \TXT(\Delta)\\ w = \pi(\tau)}}
  (S,\tau).
\end{align*}

\begin{proposition}\label{prop:toproj2algebraic}
  Let $S: \TXT(\Delta) \to \K$ be regular. Then $\pi(S):\Delta^* \to
  \K$ is an algebraic formal power series.
\end{proposition}
\begin{proof}
  Consider a WPA $\A = (\HS,\VS, \Omega, \mu, \muop,\mucl, \lambda, \gamma)$ 
  such that $\behave{\A}=S$. Let $\cX={(\HS^2 \times \{0,1\})\cup}
  (\VS^2\times\{0,1\})$. We define an algebraic system $(P_X)_{X \in
    \cX}$ as follows: For all $h_1,h_2 \in \HS$ and $v_1,v_2 \in \VS$
  we let
  \begin{align*}
    (P_{(h_1,h_2,1)},w) & = \sum_{a \in \Delta}\mu(h_1,a,h_2).a +
    \sum_{v,v' \in \VS} \sum_{s \in\Omega}\muop(h_1,(_s,v)\cdot\mucl(v',)_s,h_2).(v,v',0)\\
    (P_{(h_1,h_2,0)},w) & =
    \sum_{h_3\in \HS} (h_1,h_3,1)(h_3,h_2,1) + (h_1,h_3,1)(h_3,h_2,0)\\
    (P_{(v_1,v_2,1)},w) & = \sum_{a \in \Delta}\mu(v_1,a,v_2).a +
    \sum_{h,h' \in \HS} \sum_{s
      \in\Omega}\muop(v_1,(_s,h)\cdot\mucl(h',)_s,v_2).(h,h',0)\\
    (P_{(v_1,v_2,0)},w) & =
    \sum_{v_3\in \HS} (v_1,v_3,1)(v_3,v_2,1) + (v_1,v_3,1)(v_3,v_2,0)\\
  \end{align*}

  We claim that this algebraic system has a unique quasiregular
  solution $(S_{X})_{X\in \cX}$ which consists of algebraic formal
  power series. Indeed, if we replace the polynomial $P_{(h_1,h_2,1)}$
  by the polynomial
  \begin{align*}
    &\sum_{a \in \Delta}\mu(h_1,a,h_2).a+ \sum_{v,v' \in \VS} \sum_{s
      \in\Omega}\muop(h_1,(_s,v)\cdot\mucl(v',)_s,h_2).P_{(v,v',0)}
    \intertext{ and the polynomial $P_{(v_1,v_2,1)}$ by the
      polynomial} &\sum_{a \in \Delta}\mu(v_1,a,v_2).a+ \sum_{h,h' \in
      \HS} \sum_{s
      \in\Omega}\muop(v_1,(_s,h)\cdot\mucl(h',)_s,v_2).P_{(h,h',0)},
  \end{align*}
  we obtain an equivalent system (cf.  manipulations after
  Definition~\ref{def:algseries}) which is proper and has thus the
  unique quasiregular solution $(S_{X})_{X\in \cX}$ which consists of
  algebraic formal power series.  Let $\TXT^\circ\subseteq
  \TXT(\Delta)$ be the set of all alternating texts which are either
  singletons or $\circ$-products. Analogously let $\TXT^\bullet
  \subseteq \TXT(\Delta)$ be the set of all alternating texts which
  are either singletons or $\bullet$-products.  We will show by
  induction that we have for all $w \in \Delta^*$ with $|w| \geq 1$
  \begin{align} \label{eq:txtalge1} (S_{(h_1,h_2,1)},w) &=
    \sum_{\substack{\tau \in \TXT^\bullet \\ \pi(\tau)=w}}\sum_{r:h_1
      \stackrel{\tau}{\to}h_2} \weight_{\A}(r) \text{ and }
    \\\label{eq:txtalge2} (S_{(v_1,v_2,1)},w) &= \sum_{\substack{\tau
        \in \TXT^\circ\\ \pi(\tau)=w}}\sum_{r:v_1
      \stackrel{\tau}{\to}v_2} \weight_{\A}(r) \intertext{ as well as}
    \label{eq:txtalge3}
    (S_{(h_1,h_2,1)},w)+(S_{(h_1,h_2,0)},w) &= \sum_{\substack{\tau
        \in \TXT(\Delta) \\ \pi(\tau)=w}}\sum_{r:h_1
      \stackrel{\tau}{\to}h_2} \weight_{\A}(r) \text{ and
    }\\\label{eq:txtalge4} (S_{(v_1,v_2,1)},w)+(S_{(v_1,v_2,0)},w) &=
    \sum_{\substack{\tau \in \TXT(\Delta) \\ \pi(\tau)=w}}\sum_{r:v_1
      \stackrel{\tau}{\to}v_2} \weight_{\A}(r).
\end{align}
The result then follows immediately from the fact that algebraic
formal power series are closed under pointwise sum and scalar
multiplication.  Let $w=a$ for some $a \in \Delta$. Since the series
$S_{(h_1,h_2,1)}$ and $S_{(v_1,v_2,1)}$ are quasiregular, we obtain
that $(S_{(v_1,v_2,0)},a)=(S_{(h_1,h_2,0)},a)=0$. From this it is easy
to deduce the induction base.  Let now $|w|>1$.  Then
\begin{align*}
  &(S_{(h_1,h_2,1)},w)
  =\sum_{v,v'\in\VS}\sum_{s\in\Omega}\muop(h_1,(_s,v)\cdot\mucl(v',)_s,h_2)\cdot(S_{(v,v',0)},w)\\
  &=\sum_{v,v'\in\VS}\sum_{s\in\Omega}\muop(h_1,(_s,v)~\cdot\\*
  & \hspace{1.5em}
  \Big(\sum_{v_3\in\VS}\sum_{w=w_1w_2}(S_{(v,v_3,1)},w_1)\cdot(S_{(v_3,v',1)},w_2)+
  (S_{(v,v_3,1)},w_1)\cdot(S_{(v_3,v',0)},w_2)\Big)
  \cdot\mucl(v',)_s,h_2)\\
  &=\sum_{v,v'\in\VS}\sum_{s\in\Omega}\muop(h_1,(_s,v)~\cdot\\*
  &\hspace{3em}\Big(\sum_{v_3\in
    \VS}\sum_{w=w_1w_2}(S_{(v,v_3,1)},w_1)\cdot\big((S_{(v_3,v',1)},w_2)+
  (S_{(v_3,v',0)},w_2)\big)\Big)
  \cdot~\mucl(v',)_s,h_2)\\
  &=\sum_{v,v'\in\VS}\sum_{s\in\Omega}\muop(h_1,(_s,v)~\cdot\\*
  &\hspace{3em}\Big(\sum_{v_3\in \VS}\sum_{w=w_1w_2}
  \sum_{\substack{\tau_1 \in \TXT^\circ \\
      \pi(\tau_1)=w_1}}\sum_{r_1:v \stackrel{\tau_1}{\to}v_3}
  \weight_{\A}(r_1)\cdot \sum_{\substack{\tau_2 \in \TXT(\Delta) \\
      \pi(\tau_2)=w_2}}\sum_{r_1:v_3 \stackrel{\tau_2}{\to}v'}
  \weight_{\A}(r_1)\Big) \cdot\mucl(v',)_s,h_2) \intertext{Since
    $\TXT(\Delta)$ is the free bisemigroup, given some $w$ of length
    at least two, each $\tau \in \TXT^\bullet$ with $\pi(\tau)=w$
    decomposes uniquely into $\tau=\tau_1\bullet \tau_2$ with
    $\tau_1\in \TXT^\circ$ and $\tau_2 \in \TXT(\Delta)$. Hence we can
    continue} &= \sum_{\substack{\tau \in \TXT^\bullet \\
      \pi(\tau)=w}}\sum_{r:h_1 \stackrel{\tau}{\to}h_2}
  \weight_{\A}(r).
\end{align*}
Analogously we get Equation~(\ref{eq:txtalge2}). Similarly we get:
\begin{align*}
  (S&_{(h_1,h_2,1)},w)+(S_{(h_1,h_2,0)},w)
  =\\
  =&\hspace{-.5em}\sum_{\substack{\tau \in \TXT^\bullet \\
      \pi(\tau)=w}}\sum_{r:h_1 \stackrel{\tau}{\to}h_2}\hspace{-.5em}
  \weight_{\A}(r)~+
  \sum_{h_3\in\HS}\sum_{w=w_1w_2}(S_{(h_1,h_3,1)},w_1)\cdot\big((S_{(h_3,h_2,1)},w_2)
  +
  (S_{(h_3,h_2,0)},w_2)\big)\\
  =&\sum_{\substack{\tau \in \TXT^\bullet \\
      \pi(\tau)=w}}\sum_{r:h_1
    \stackrel{\tau}{\to}h_2} \weight_{\A}(r)~+ \\*
  &\hspace{7em}\sum_{h_3\in \HS}\sum_{w=w_1w_2}
  \sum_{\substack{\tau_1 \in \TXT^\bullet \\
      \pi(\tau_1)=w_1}}\sum_{r_1:h_1 \stackrel{\tau_1}{\to}h_3}
  \weight_{\A}(r_1)\cdot \sum_{\substack{\tau_2 \in \TXT(\Delta) \\
      \pi(\tau_2)=w_2}}\sum_{r_1:h_3
    \stackrel{\tau_2}{\to}h_2} \weight_{\A}(r_1)\\
  =&\sum_{\substack{\tau \in \TXT^\bullet \\
      \pi(\tau)=w}}\sum_{r:h_1 \stackrel{\tau}{\to}h_2}
  \weight_{\A}(r)~+ \sum_{\substack{\tau \in \TXT^\circ \\
      \pi(\tau)=w}}\sum_{r:h_1
    \stackrel{\tau}{\to}h_2} \weight_{\A}(r)\\
  =& \sum_{\substack{\tau \in \TXT(\Delta) \\
      \pi(\tau)=w}}\sum_{r:h_1 \stackrel{\tau}{\to}h_2}
  \weight_{\A}(r).
\end{align*}
Again Equation~(\ref{eq:txtalge4}) can be shown analogously, which
concludes the proof. \eop
\end{proof}

Now we get our second characterization of algebraic formal power
series. For this, we proceed as follows: Let $\phi$ be a weighted
second-order formula over words containing, apart from a single
$2$-ary relation variable $\leq_2$, only $1$-ary relation
variables. In other words, let $\phi\in \MSO(\K,\sigtxt{\Delta})$.
Let $\Free(\phi) \subseteq \V$, $w = ([n],\leq_1,\lambda) \in
\Delta^*$ and $\gamma$ a $(\V,w)$-assignment.  We define the semantics
$\seman{\exists \leq_2.  \phi}^{\text{tdo}}: \Delta^* \to \K$ by
letting
$$(\seman{\exists \leq_2.  \phi}^{\text{tdo}},(w,\gamma)) =
\sum\limits_{\leq_2 \in \TDO_{n}} (\seman{\phi},(([n],\leq_1,\leq_2,\lambda),\gamma)).$$

\begin{theorem}\label{thm:algebraiccharac-txt}
  Let $\K$ be a commutative semiring and let $S: \Delta^* \to \K$ be a
  formal power series. Then the following are equivalent:
  \begin{enumerate}[\em(1)]
  \item $S$ is an algebraic formal power series. 
  \item $S = \pi(R)$ for some regular $R:\TXT(\Delta) \to \K$. 
  \item There is a sentence $\phi \in \sRFO(\K,\sigtxt{\Delta})$ such
    that $\seman{\exists \leq_2. \phi}^{\text{tdo}} =S$.
  \end{enumerate}
\end{theorem}
\begin{proof}
  \noindent{(1) $\Rightarrow$ (3)}.\enspace Let $S:\Delta^*\to\K$ be an algebraic
  formal power series. By Theorem~\ref{thm:algebraiccharac} there is
  an $\sRFO(\K)$ sentence over nested words such that $\seman{\exists
    \nu.  \phi}^{\text{nest}}=S$.  By
  Corollary~\ref{cor:nwembedd-phi-1def} the partial function
  $\Phi_\circ^{-1}$ is $\FO$-definable without parameter. Similar to
  Proposition~\ref{prop:deftrans-weightdef} one can show that there is
  thus an $\sRFO(\K)$ sentence $\phi'$ over texts such that
  $\Phi_\circ(\seman{\phi})=\seman{\phi'}$. Now we can calculate using
  observations~(\ref{enum:phi-observ1}) and~(\ref{enum:phi-observ2})
  before Lemma~\ref{lem:trans-evennestdep} as follows.
  \begin{align*}
    (\seman{\exists \nu. \phi}^{\text{nest}},w) &= \sum_{\substack{nw
        \in \NW(\Delta) \\ \pi(nw)=w}}(\seman{\phi},nw)
    = \sum_{\substack{nw \in \NW(\Delta)\\ \pi(\Phi_\circ(nw))=w}}(\seman{\phi},nw)\\
    &= \sum_{\substack{\tau \in \Phi_\circ(\NW(\Delta))\\
        \pi(\tau)=w}}(\seman{\phi},\Phi_\circ^{-1}(\tau))
    = \sum_{\substack{\tau \in \Phi_\circ(\NW(\Delta))\\
        \pi(\tau)=w}}(\seman{\phi'},\tau) =(\seman{\exists
      \leq_2. \phi'}^{\text{tdo}},w).
  \end{align*}

  \noindent{(3) $\Rightarrow$ (2)}.\enspace Follows from
  Theorem~\ref{thm:txt-srmso} and the definition of $\pi$.

  \noindent{(2) $\Rightarrow$ (1)}.\enspace This is
  Proposition~\ref{prop:toproj2algebraic}.\eop
\end{proof}

\section{Concluding Remarks and Future Work} 
We introduced a quantitative automaton model and a quantitative logic
for nested words and showed that they are equally expressive. This
generalizes the logical characterization of the unweighted case as
given in \cite{AluMad06}. Moreover, we established a new connection
between nested words and alternating texts.  Applying the result, we
obtained a characterization of algebraic formal power series in terms
of weighted logics.  Presumably, the logical characterization of
regular nested word series could also be obtained by structural
induction. However, the connection between alternating texts and
nested words enabled us to also obtain a second characterization of
algebraic formal power series. Note that even though the
characterizations of algebraic formal power series are generalizations
of the results of~\cite{Lauetal94} to a weighted setting, in contrast
to the latter paper we gave a different proof using this connection as
well as (weighted) nested word automata and (weighted) parenthesizing
automata. Also note that weighted nested word automata and weighted parenthesizing
automata were characterized algebraically in~\cite{Mat09a}. 

Let us remark that regular formal power series also fall into the
pattern of our characterizations (Theorem~\ref{thm:algebraiccharac}
and Theorem~\ref{thm:algebraiccharac-txt}) of algebraic formal power
series. In fact, Thomas showed that a single existential monadic
second-order quantifier suffices to characterize finite
automata~\cite[Theorem~5.2]{Tho82}. That is, in the pattern of the
last results we can formulate that $L\subseteq \Delta^*$ is regular
iff $L=\seman{\exists M.\phi}^{\text{set}}$ for some first-order
formula $\phi$ (where $\seman{\exists M.\phi}^{\text{set}}$ means that
we sum over all subsets $M$ of the domain of a given structure).  Let
us explain the idea of the proof with an example.  Given an automaton
$\A$ with set of states $Q=\{0,1\}^k$ for $L$ and some word $a_1\ldots
a_{2k}\in L$, the idea is to think of the interpretation of $M$ as a
word $u_1u_2$ where $u_1,u_2\in \{0,1\}^k=Q$ and to express by $\phi$
that $u_1$ is an initial state, that there is a run from $u_1$ to
$u_2$ on $a_1\ldots a_k$ and that there is a run from $u_2$ into a
final state on $a_{k+1}\ldots a_{2k}$.  Alternatively, one can prove
the result similarly to Proposition~\ref{prop:algseraredef} where one
starts with a right-regular system and applies a similar
transformation. Then a set $M$ suffices to encode a derivation tree
since any inner node has at most one non-terminal child whose position
is collected in $M$.  In any way, it is not hard to see that the proof
can be adapted to a weighted setting. So, also in the weighted case we
can restrict ourselves to a single existential monadic second-order
quantifier.

Following these pattern it might be interesting to further
investigate whether other important classes of formal power series can be
characterized in this manner. Again, the work of Lautemann, Schwentick
and Th{\'e}rien~\cite{Lauetal94} can be used as a starting point where
the so-called $k$-linear languages were considered. 

\section*{Acknowledgments.} 
The author thanks Manfred Droste and Andreas Maletti for helpful
comments, Dietrich Kuske for pointing him to
\cite{Lauetal94}. Moreover, he thanks the anonymous referees of this
journal version and the referees of the conference version.  Their
careful reading and their remarks resulted in substantial
improvements.

\end{document}